%% file: 46th.tex
\documentclass[11pt]{article}
\usepackage{epsfig,amsmath,color,amssymb,euscript,epstopdf}
\usepackage{latexsym,amsfonts}
\renewcommand{\baselinestretch}{1.5}
\makeatletter
\input{QYalias}

\newcommand{\beqn}{\begin{eqnarray}}             
\newcommand{\eeqn}{\end{eqnarray}}               
\newcommand{\beq}{\begin{eqnarray*}}             
\newcommand{\eeq}{\end{eqnarray*}}
\newcommand{\red}[1]{\textcolor{red}{#1}}

\newcommand{\nn}{\nonumber}

\newcommand{\ignore}[1]{}{}
    {\setcounter{araenumcount}{0}%
    \begin{list}{\arabic{araenumcount}.}{\usecounter{araenumcount}}}%
    {\end{list}}
    {\setcounter{roenumcount}{0}%
    \begin{list}{(\Roman{roenumcount})}{\usecounter{roenumcount}}}%
    {\end{list}}
    {\setcounter{romenumcount}{0}%
    \begin{list}{(\roman{romenumcount})}{\usecounter{romenumcount}}}%
    {\end{list}}

\newtheorem{remark}{Remark} 

\begin{document}

\title{\large \bf Krigings Over Space and Time Based on  Latent\\ Low-Dimensional
Structures\footnote{Partially supported by
National Statistical Research Project of China 2015LY77 and
NSFC grants 11571080, 11571081, 71531006 (DH),
by EPSRC grant EP/L01226X/1 (QY), and by
NSFC grants 11371318 (RZ).}}
\author{\normalsize Da Huang$^{\dagger}$ \qquad Qiwei Yao$^{\ddagger}$
\qquad Rongmao Zhang$^\star$\\[-1ex]
\small
$^{\dagger}$School of Management, Fudan University, Shanghai, 200433,  China\\[-1ex]
\small
$^{\ddagger}$Department of Statistics, London School of Economics, London, WC2A 2AE, U.K.\\[-1ex]
\small
 $^\star$School of Mathematics, Zhejiang University, Hangzhou, 310058, China\\[-1ex]
\small
dahuang@fudan.edu.cn \quad q.yao@lse.ac.uk \quad rmzhang@zju.edu.cn
}

\date{}

\maketitle

\begin{abstract}
We propose a new approach to represent nonparametrically the linear dependence structure of a spatio-temporal process in terms of latent common factors.  Though it is formally similar to the existing reduced rank approximation methods (Section 7.1.3 of Cressie and Wikle, 2011), the fundamental difference is that the low-dimensional structure is completely unknown in our setting, which is learned from the data collected irregularly over space but regularly over time.  Furthermore a graph Laplacian is incorporated in the learning in order to take the advantage of the continuity over space, and a new aggregation method via randomly partitioning space is introduced to improve the efficiency.  We do not impose any stationarity conditions over space either, as the learning is facilitated by the stationarity in time.  Krigings over space and time are carried out based on the learned low-dimensional structure, which is scalable to the cases when the data are taken over a large number of locations and/or over a long time period.  Asymptotic properties of the proposed methods are established.  Illustration with both simulated and real data sets is also reported.
\end{abstract}

\bigskip


\section{Introduction}
\setcounter{equation}{0}

Kriging, referring to the spatial best linear prediction, is named by Matheron after South
African mining engineer Daniel Krige.
The key step in kriging is to identify and to estimate the covariance structure.
The early applications of kriging are
typically based on some parametric models for spatial covariance functions.
See Section 4.1 of Cressie and Wikle (2011) and references within.
However fitting those parametric covariance models to large spatial or spatio-temporal datasets is
conceptually indefensible (Hall, Fisher and Hoffmann, 1994).
It also poses serious computational challenges. For example, a spatial kriging
with observations from $p$ locations involves inverting a $p\times p$ covariance
matrix, which typically requires $O(p^3)$ operations with $O(p^2)$ memory.
One attractive approach to overcome the computational burden is to introduce
reduced rank approximations for the underlying processes. Methods in
this category include Higdon (2002) using kernel convolutions,  Wikle and Cressie (1999),
Kammann and Wand (2003) and Cressie and Johannesson (2008) using low rank basis functions
(see also Section 7.1.3 of Cressie and Wikle, 2011), Banerjee \etal (2008) and
Finley \etal (2009) using predictive processes, and Tzeng and Huang (2018)
using thin-plate splines. However as pointed out by Stein (2008),
the reduced rank approximations often fail to capture small-scale correlation structure
accurately. An alternative approach is to seek sparse approximations for
covariance functions, see, e.g., Gneiting (2002) using compactly supported
covariance functions, and Kaufman, Schervish and Nychka (2008) proposing
a tempering method by setting the covariances to 0 between any two locations
with the distances beyond a threshold. Obviously these approaches
miss the correlations among the locations which are distantly apart
from each other. Combining together both the ideas of
reducing rank and the tempering,
Sang and Huang (2012) and  Zhang, Sang and Huang (2015) proposed a so-called full scale
approximation method for large spatial and spatio-temporal datasets.

In this paper we propose a
new nonparametric approach to represent the linear dependence structure
of a spatio-temporal process. 
Different from all the methods
stated above, we impose neither any distributional assumptions on the
underlying process nor any parametric forms on its covariance function.
Under the setting that the observations are taken irregularly over
space but regularly in time, we recover the linear dependent structure based on
a latent factor representation. No stationarity conditions are imposed
over space, though the stationary in time is assumed. Formally
our latent factor model is a reduced rank representation. However both
the factor process and the factor loadings are completely unknown. This
is a marked difference from the aforementioned reduced rank
approximation methods. The motivation for our approach is to learn the linear dynamic
structure across both space and time directly from data with little subjective
assumptions.  It captures the dependence across the locations over all
distances automatically.

The latent factors and the corresponding loadings are estimated via an eigenanalysis.
However it differs from the eigenanalysis for estimating latent factors
for multiple time series (cf. Lam and Yao, 2012, and the references within) in
at least three
aspects. First, we extract the information from the dependence across
different locations instead of over time: the whole observations are
divided into two sets according to their locations, the
estimation boils down to the singular value decomposition (SVD) of the
spatial covariance matrix of two data sets. One advantage of this
approach is that it is free from the impact of the `nugget effect' in the
sense that we do not need to estimate the variances of, for example,
measurement errors in order to recover the latent dependence structure.
Secondly, we propose new aggregation via randomly partitioning the
observations over space to improves the original estimation.
This also overcomes the arbitrariness in dividing data in the eigenanalysis.
The aggregation proposed is in the spirit of the Bagging of Breiman  (1996),
though random partitioning instead of bootstraping is
used in our approach.
Thirdly, we incorporate a graph Laplician (Hastie \etal, 2009, pp.545) into the
eigenanalysis to take the advantage of the continuity over space, leading
to further improvement in both estimation and kriging.

The number of latent factors is typically small or at least much smaller
than the number of
locations on which the data are recorded. Consequently the krigings can be
performed via only inverting matrices of the size equal to the number of factors.
This is particularly
appealing when dealing with large datasets. However the SVD for estimating the latent
factor structure requires $O(p^3)$ operation. Nevertheless the nonparametric nature
makes our approach easily scalable to large datasets.
See Section \ref{sec33} below.

It is worth pointing out that our approach is designed for analyzing
spatio-temporal data or pure spatial data but with repeated observations.
With the advancement of information technology,
large amount of data are collected routinely over space and time nowadays.
The  surge of the development of statistical methods and theory for modelling and forecasting
spatio-temporal processes includes, among others,
Smith, Kolenikov and Cox (2003),
 Jun and Stein (2007), Li, Genton and Sherman (2007), Katzfuss and
Cressie (2011),  Castruccio and Stein (2013), Guinness and Stein (2013),
Zhu, Fan and Kong (2014), Zhang, Sang and Huang (2015), and Wang and Huang (2017).
See also the monograph Cressie and Wilkle (2011).
In addition to the methods based on low-dimensional covariance
structures, the dynamic approach
which, typically, specifies the standard Gaussian autoregressive model of
order 1 (i.e. AR(1)), coupled with MCMC computation has gained popularity in
modelling large spatio-temporal data. Cressie, Shi and Kang (2010) 	
assumed a Gaussian AR(1) model for a low-dimensional latent process and developed
a full scale Kalman filter in the context of spatio-temporal modelling.
See also Chapter 7 of Cressie and Wilkle (2011).

The rest of the paper is organized as follows. We specify the latent factor structure for
a spatio-temporal process in Section \ref{sec2}. The newly proposed estimation methods
are spelt out in Section \ref{sec3}. The kriging over space and time is
presented in Section \ref{sec4},
 in which we also state how to handle missing values.  The asymptotic
results for the proposed estimation and kriging methods are
developed in Section \ref{sec5}.
Illustration with both simulated and real data is reported in Section \ref{sec6}.
 Technical proofs are relegated to the Appendix in a supplementary file.

\section{Models} \label{sec2}
\subsection{Setting}
Consider spatio-temporal process
\begin{equation} \label{b1}
y_t (\bs) = \bz_t(\bs)'\bbeta(\bs)+  \xi_t(\bs) +  \ve_t(\bs), \quad t=0, \pm
1, \pm 2, \cdots, \; \bs \in \calS
\subset \RR^2,
\end{equation}
where $\bz_t(\bs)$ is an $m \times 1$ observable covariant vector, $\bbeta(\bs)$ is
a unknown parameter vector,
 $\ve_t(\bs) $ is unobservable and  represents the so-called nugget effect (in space) in the sense that
\begin{equation} \label{b2}
E\{ \ve_t(\bs)\} =0, \quad \var\{ \ve_t(\bs)\}= \sigma(\bs)^2, \quad
\cov\{ \ve_{t_1}(\bu), \ve_{t_2}(\bv) \} =0 \; \; \forall \;  (t_1, \bu) \ne (t_2, \bv),
\end{equation}
$ \xi_t(\bs)$ is a latent spatio-temporal process satisfying the conditions
\begin{equation} \label{b3}
E\{ \xi_t(\bs)\} =0, \qquad \cov\{ \xi_{t_1}(\bu), \xi_{t_2}(\bv) \} =
\Sigma_{|t_1-t_2|}(\bu, \bv).
\end{equation}
Consequently, $y_t(\bs) - \bz_t(\bs)'\bbeta(\bs)$ is (weakly) stationary in time $t$,
$E\{ y_t(\bs) -\bz_t(\bs)'\bbeta(\bs)\}=0$, and
\begin{equation}
 \cov\{ y_{t_1}(\bu)- \bz_{t_1}(\bu)'\bbeta(\bu),\; y_{t_2}(\bv)- \bz_{t_2}(\bv)'\bbeta(\bv) \}
=
\Sigma_{|t_1-t_2|}(\bu, \bv)
+ \sigma(\bu)^2 I\{ (t_1, \bu) = (t_2, \bv) \}.
\label{b5n}
\end{equation}
Furthermore we assume that $\Sigma_t(\bu, \bv)$
is continuous in $\bu$ and $\bv$.

Model (\ref{b1}) does not impose any stationarity conditions
over space. However it requires
that $y_t(\cdot) -  \bz_t(\cdot) ' \bbeta(\cdot)$  is second order stationary in time $t$,
which enables the learning of the dependence across different locations
and times. In practice the data often show some trends and seasonal patterns in time. 
The existing detrend and deseasonality methods in time series analysis can be applied to 
make data stationary in time.


\subsection{A finite dimensional representation for $\xi_t(\bs)$}

Let $L_2(\calS)$ be the Hilbert space consisting of all the square integrable functions
defined on $\calS$ equipped with the inner product
\begin{equation} \label{b5}
\inner{f}{g} = \int_{\calS} f(\bs) g(\bs) d\bs, \qquad f, g \in L_2(\calS).
\end{equation}
We assume that the latent process $\xi_t(\bs)$ admits a finite-dimensional structure:
\begin{equation} \label{b4}
\xi_t (\bs) = \sum_{j=1}^d a_j(\bs) x_{tj},
\end{equation}
where $a_1(\cdot), \cdots, a_d(\cdot)$ are deterministic and linear independent
functions
(i.e. none of them can be written as a linear combination of the others)
in the Hilbert space $L_2(\calS)$, and $x_{t1}, \cdots,
x_{td}$ are $d$ latent time series.
Obviously $a_1(\cdot), \cdots, a_d(\cdot)$ (as well as $x_{t1}, \cdots, x_{td}$) are
not uniquely defined by (\ref{b4}), as they can be replaced by any of their non-degenerate linear
transformations.
 There is no loss
of generality in assuming that $a_1(\cdot), \cdots, a_d(\cdot)$ are
orthonormal in the sense
that
\begin{equation} \label{b6}
\inner{a_i}{a_j} = I(i=j),
\end{equation}
as any set of linear independent functions in a Hilbert space can be standardized to
this effect.
Let $\bx_t = (x_{t1}, \cdots, x_{td})'$. It follows from (\ref{b3}) that $\bx_t$ is a $d$-variant
stationary time series with mean $\bf0$, and
\begin{equation} \label{b10n}
\Sigma_0(\bu, \bv) = \cov\{ \xi_{t}(\bu), \xi_{t}(\bv) \} =
\sum_{i,j=1}^d a_i(\bu) a_j(\bv) \sigma_{ij},
\end{equation}
where $\sigma_{ij}$ is the $(i,j)$-th element of $\var(\bx_{t})$.
Let
\begin{equation} \label{b10}
\Sigma_0 \circ f (\bs) = \int_\calS \Sigma_0(\bs, \bu) f(\bu) d\bu, \qquad f \in L_2(\calS).
\end{equation}
Then $\Sigma_0$ is a non-negative definite operator defined on $L_2(\calS)$.
See Appendix A of Bathia \etal (2010) for some basic facts on the operators in Hilbert spaces.
It follows from Mercer's theorem (Mercer 1909) that $\Sigma_0$ admits the
spectral decomposition
\begin{equation} \label{b8}
\Sigma_0(\bu, \bv) = \sum_{j=1}^d \lambda_j \varphi_j(\bu) \varphi_j(\bv),
\end{equation}
where $\la_1 \ge \cdots \ge \la_d >0$ are the $d$ positive eigenvalues of $\Sigma_0(\bu, \bv)$, and
$\varphi_1, \cdots, \varphi_d \in L_2(\calS)$ are the corresponding eigenfunctions, i.e.
\begin{equation} \label{b9}
\Sigma_0\circ \varphi_j(\bs) = \int_\calS \Sigma_0(\bs, \bu) \varphi_j(\bu) d \bu = \la_j \varphi_j(\bs).
\end{equation}
See Proposition~\ref{prop1}  below.

\begin{proposition} \label{prop1}
Let rank$(\var(\bx_t))=d$. Then the following assertions hold.
\begin{itemize}
\item[(i)]
$\Sigma_0$ defined in (\ref{b10}) has exactly $d$ positive eigenvalues.
\item[(ii)]
The $d$ corresponding orthonormal eigenfunctions can be expressed as
\[
\varphi_i(\bs) = \sum_{j=1}^d \gamma_{ij} a_j(\bs), \qquad i=1, \cdots, d,
\]
where $\bgamma_i \equiv (\gamma_{i1}, \cdots, \gamma_{id})'$, $i=1, \cdots, d$, are $d$
orthonormal eigenvectors of matrix $ \var(\bx_t) $.
\end{itemize}
\end{proposition}

\askip

The above proposition shows that the finite-dimensional structure (\ref{b4}) can be
identified via the covariance functions of $\xi_t(\bs)$, though the representation of (\ref{b4}) itself
is not unique.  Note that the linear space spanned by
the eigenfunctions $\varphi_1(\cdot), \cdots, \varphi_d(\cdot)$ is called the kernel reproducing Hilbert space
(KRHS) by $\Sigma_0(\cdot, \, \cdot)$, and $\{ a_j(\cdot) \}$ and $\{
\varphi_j(\cdot)\}$ are two orthonormal bases for this KRHS.
Furthermore any orthonormal basis of this KRHS can be taken as $a_1(\cdot), \cdots, a_d(\cdot)$.
In Section \ref{sec3} below, the estimation for $a_1(\cdot), \cdots, a_d(\cdot)$
will be constructed in this spirit.

\section{Estimation}
\label{sec3}

Let $\{ (y_t(\bs_i), \bz_t(\bs_i) ), \; i=1, \cdots, p, \; t=1, \cdots, n\}$ be the available observations
over space and time, where $\calS_o \equiv \{\bs_1, \cdots, \bs_p \}
 \subset \calS $ are typically irregularly spaced.
The total number of observations is $n\cdot p$.

\subsection{Estimation for finite dimensional representations of $\xi_t(\bs)$}
\label{sec31}
To simplify the notation, we first consider a special case $\bbeta(\bs)
\equiv 0$ in (\ref{b1}) in Sections~\ref{sec31} \& \ref{sec32}.
Section \ref{sec34} below considers the least squares regression
estimation for $\bbeta(\bs)$. Then the procedures
describe in  Sections~\ref{sec31} \& \ref{sec32} still apply if $\{
y_t(\bs_i)\}$ are replaced by the
residuals from the regression estimation.

Now under (\ref{b4}),
\begin{equation}\label{c2}
y_t(\bs) = \xi_t(\bs)+ \ve_t(\bs)= \sum_{j=1}^d a_j(\bs) x_{tj} + \ve_t(\bs).
\end{equation}
 To exclude
nugget effect in our estimation, we divide $p$ locations $\bs_1, \cdots, \bs_p$ into
two sets $\calS_1$ and $\calS_2$ with, respectively, $p_1$ and $p_2$ elements, and
$p_1+p_2=p$.
Let $\by_{t, i}$ be a vector consisting of $y_t(\bs)$ with $ \bs \in \calS_i$, $i=1, 2$.
Then $\by_{t, 1}, \by_{t, 2}$ are two vectors with lengths $p_1$ and $p_2$ respectively.
Denoted by $\bxi_{t,1}, \, \bxi_{t, 2}$ the corresponding vectors consisting of $\xi_t(\cdot)$.
It follows from (\ref{c2}) that
\begin{equation} \label{f1}
\by_{t,1} = \bxi_{t,1} + \bve_{t,1}= \bA_1 \bx_t + \bve_{t,1}, \qquad
\by_{t,2} = \bxi_{t, 2} + \bve_{t,2}= \bA_2 \bx_t + \bve_{t,2},
\end{equation}
where $\bA_i$ is a $p_i \times d$ matrix, its rows consist of the coefficients
$a_j(\cdot)$ on the RHS of (\ref{c2}), and $\bve_{t,i}$ consists of $\ve_t(\bs)$
with $\bs \in \calS_i$. There is no loss of generality in assuming
$\bA_1' \bA_1 =\bI_d$. This can be achieved by performing  an orthogonal-triangular (QR)
decomposition $\bA_1 = \bGamma \bR$, and replacing $(\bA_1, \bx_t)$ by $(\bGamma, \bR\bx_t)$
in the first equation in (\ref{f1}). Note $\calM(\bA_1) = \calM(\bGamma)$, where
$\calM(\bA)$ denotes the linear space spanned by the columns of matrix $\bA$.
Thus $\calM(\bA_1)$ does not change from imposing the condition $\bA_1' \bA_1 =\bI_d$.
 Similar we may also assume $\bA_2' \bA_2 =\bI_d$, which however implies that
$\bx_t$ in the second equation in (\ref{f1}) is unlikely to be the same as that
in the first equation. Hence we may re-write (\ref{f1}) as
\begin{equation} \label{f2}
\by_{t,1} =
\bA_1 \bx_t + \bve_{t,1}, \qquad
\by_{t,2} =
\bA_2 \bx_t^\star + \bve_{t,2},
\end{equation}
where $\bA_1'\bA_1 = \bA_2' \bA_2= \bI_d$, $ \bx_t^\star = \bQ \bx_t $, and
$\bQ$ is an invertible $d \times d$ matrix. Note that $(\bA_1, \bx_t)$
and $(\bA_2, \bx_t^\star)$ are still not uniquely defined in (\ref{f2}), as they can be
replaced, respectively, by  $(\bA_1 \bGamma_1, \bGamma_1'\bx_t)$
and $(\bA_2 \bGamma_2,  \bGamma_2'\bx_t^\star)$ for any $d \times d$ orthogonal matrices
$\bGamma_1$ and $\bGamma_2$.  However $\calM(\bA_1)$ and $\calM(\bA_2)$ are uniquely
defined by (\ref{f2}).

Since $\by_{t,1}$ and $\by_{t,2}$ have no common
elements, it follows from (\ref{c2}) and (\ref{b2}) that
\begin{equation} \label{f5n}
\bSigma \equiv \cov(\by_{t,1}, \by_{t,2}) = \bA_1 \cov(\bx_t, \bx_t^{\star}) \bA_2'.
\end{equation}
Note that $\cov(\bx_t, \bx_t^{\star}) = \var(\bx_t) \bQ$.
When $p \gg d$, it is reasonable to assume that rank$(\bSigma) =
{\rm rank}\{\cov(\bx_t, \bx_t^\star)\} ={\rm rank}\{\var(\bx_t)\} =d$. Let
\begin{equation} \label{f5}
\bSigma \bSigma ' = \bA_1 \cov(\bx_t, \bx_t^\star) \cov(\bx_t^\star, \bx_t)\bA_1', \qquad
\bSigma ' \bSigma = \bA_2 \cov(\bx_t^\star, \bx_t) \cov(\bx_t, \bx_t^\star) \bA_2'.
\end{equation}
Then these two matrices share the same $d$ positive eigenvalues, and
$\bSigma \bSigma ' \bb =0$ for any vector $\bb$ perpendicular to $\calM(\bA_1)$.
Therefore, the $d$ orthonormal eigenvectors of matrix  $\bSigma \bSigma
'$ corresponding to its $d$ positive eigenvalues
can be taken as the columns of $\bA_1$. Similarly
the $d$ orthonormal eigenvectors of matrix $\bSigma '\bSigma $
corresponding to its $d$ positive eigenvalues can be taken as the columns of $\bA_2$.
We construct the estimators for $\bA_1, \, \bA_2$ based on this observation.

Let $\wh \bSigma$ be the sample covariance of $\by_{t,1}$ and $\by_{t,2}$, i.e.
\begin{equation} \label{f3}
\wh \bSigma = {1 \over n} \sum_{t=1}^n( \by_{t,1} - \bar \by_1) (\by_{t,2} - \bar \by_2)',
\end{equation}
where $\bar \by_i = n^{-1} \sum_t \by_{t,i}$. Let $\wh \la_1 \ge \wh \la_2 \ge \cdots $
be the eigenvalues of $\wh \bSigma\wh \bSigma'$.
A natural estimator for $d$ is defined as
\begin{equation} \label{c5}
\wh d = \max_{1 \le j < p_*} \wh \la_j \big/ \wh \la_{j+1},
\end{equation}
where $p_* \ll \min( p_1, \, p_2)$  is a prespecified integer (\eg $p_* =
\min( p_1, \, p_2)/2$). This estimation method is
based on the fact that
$\la_j/\la_{j+1}$ are positive and finite constants for $j=1, \cdots, d-1$, and
$\la_d/\la_{d+1} = \infty$. However $\la_j/\la_{j+1}$ is asymptotically
`$0/0$' for $j=d+1, \cdots, p-1$.
In practice, we mitigate this difficulty by
comparing the ratios for $j < p_* \ll \min(p_1, p_2)$.
Asymptotic properties of the ratio estimators under different settings have been established
in, e.g. Lam and Yao (2012), Chang \etal (2015), and Zhang \etal (2018). The (fine)
finite sample  performance of the ratio estimators are also reported in those papers.

Consequently the $\wh d$ orthonormal eigenvectors of $\wh \bSigma\wh
\bSigma'$ (or $\wh \bSigma'\wh \bSigma$), corresponding to the eigenvalues
$\wh \la_1, \cdots, \wh \la_{\wh d}$,  can be taken as the estimated columns of $\bA_1$
(or $\bA_2$).
However such an estimator ignores the fact that
$\xi_t(\cdot)$ is continuous over the set $\calS$, which should be taken into
account to improve the estimation.  To achieve this, denoted by $\bs_1^1, \cdots, \bs_{p_1}^1$	
the $p_1$ locations in $\calS_1$ arranged according to the order such that the $j$-th component of
$\by_{t,1}$ is the observation taken at the location $\bs^1_j$.
We define a graph Laplacian $\bL \equiv \bG - \bW$, where
$\bW=(w_{ij})$ is a weight matrix with $w_{ii}=0$ and, e.g. $w_{ij}= 1
/(1+ \| \bs_i^1 - \bs_j^1\|)$
($\|\cdot\|$ denotes the Euclidean norm)
for $i\ne j$, and $\bG=(g_{ij})$ with $g_{ii} = \sum_j w_{ij}$ and $g_{ij}=0$ for all
$i\ne j$.
Then it holds that for any column vector $\ba=(a_1, \cdots, a_p)'$,
\[
\ba' \bL \ba = \sum_{i=1}^p g_{ii} a_i^2 - \sum_{i,j=1}^p w_{ij} a_i a_j
= {1 \over 2} \sum_{i,j=1}^p w_{ij} (a_i - a_j)^2.
\]
See, e.g.,  Hastie, Tibshirani and Friedman (2009, pp.545). By requiring
$\ba' \bL \ba \le c_0$ for some small positive constant $c_0$, the
components of $\ba$ at the
nearby locations will be close with each other.
Hence the columns of $\bA_1$
are obtained by solving the following optimization problem:
\[
\wh \bgamma_1 =\arg \max_{ \bgamma} \bgamma' \wh \bSigma\wh
\bSigma' \bgamma \quad {\rm subject \; to} \;\; \| \bgamma\|=1 \;\;
{\rm and } \; \; \bgamma' \bL \bgamma \le c_0,
\]
and for $j=2, \cdots, \wh d$,
\[
\wh \bgamma_j = \arg \max_{ \bgamma} \bgamma' \wh \bSigma\wh
\bSigma' \bgamma \quad {\rm subject \; to} \;\; \| \bgamma\|=1, \;\; \bgamma' \wh \bgamma_i=0 \;
{\rm for\;} 1\le i <j, \;\;
{\rm and } \; \; \bgamma' \bL \bgamma \le c_0.	
\] 	
The above constrained optimization problem can be recast as an eigen-problem
for the symmetric (but not necessarily non-negative definite) matrix $\wh
\bSigma \wh \bSigma' - \tau \bL$ stated below,
where $\tau>0$  controls the
penalty according to $\bL$.
\begin{quote}
{\sl
Find the orthonormal eigenvectors $ \wh \bgamma_1, \cdots, \wh \bgamma_{\wh d}$ of $\wh
\bSigma \wh \bSigma' - \tau \bL$ corresponding to its $\wh d$ \linebreak largest eigenvalues.
}
\end{quote}
Denote the resulting estimator for the loading matrix $\bA_1$
by
\begin{equation} \label{f4}
\wh \bA_1 = (\wh \bgamma_1, \cdots, \wh \bgamma_{p_1}).
\end{equation}
The estimator for $\bA_2$, denoted by $\wh \bA_2$, is constructed in the same manner.	

By (\ref{f1}), the estimators for the two different representations of the latent processes
are defined as
\begin{equation} \label{f6}
\wh \bx_t = \wh \bA_1' \by_{t,1}, \qquad \wh \bx_t^\star = \wh \bA_2' \by_{t,2}.
\end{equation}
Consequently,
\begin{equation} \label{f7}
\wh \bxi_{t,1} = \wh \bA_1 \wh \bx_{t} = \wh \bA_1\wh \bA_1' \by_{t,1}, \qquad
\wh \bxi_{t,2} = \wh \bA_2 \wh \bx_{t}^\star = \wh \bA_2\wh \bA_2' \by_{t,2}.
\end{equation}
See also (\ref{f1}).

\begin{remark} \label{remark1}
(i) The assumption that matrix $\bSigma $ in (\ref{f5n}) has rank $d$ implies
that all the latent
factors are spatially correlated; see (\ref{c2}). In the {\sl unlikely} scenarios that some
latent factors are only serially correlated but spatially uncorrelated,
we should include autocovariance matrices in the estimation (Lam and Yao,
2012). To this end, let 
\[
\wh \bSigma_i(k) = {1 \over n} \sum_{t=1}^{n-k} (\by_{t+k,i} - \bar \by_i)
(\by_{t,i} - \bar \by_i)',
\quad 
\wh \bSigma_{12}(k) = {1 \over n} \, \sum_{t=\max\{1, -k\}}^{\min\{n-k, n\}} (\by_{t+k,1} - \bar \by_1) 
(\by_{t,2} - \bar \by_2)'.
\]
Assume $p_1=p_2$ for simplicity. Put
\[
\bM_1=\wh \bSigma \wh \bSigma' + \sum_{j=1}^{k_0}\big\{ \wh \bSigma_1(j) \wh \bSigma_1(j)'
+ \wh \bSigma_{12}(j) \wh \bSigma_{12}(j)' +
\wh \bSigma_{12}(-j) \wh \bSigma_{12}(-j)' \},
\] 
\[
\bM_2=\wh \bSigma' \wh \bSigma + \sum_{j=1}^{k_0}\big\{ \wh \bSigma_2(j) \wh \bSigma_2(j)'
+ \wh \bSigma_{12}(j)' \wh \bSigma_{12}(j) +
\wh \bSigma_{12}(-j)' \wh \bSigma_{12}(-j) \},
\] 
where $\wh \bSigma$ is defined in (\ref{f3}), and $k_0\ge 1$ is an integer. Then 
we replace $\wh \bSigma \wh \bSigma'$ by $\bM_1$ for computing $\wh \bA_1$ in (\ref{f4}),
and replace $\wh \bSigma' \wh \bSigma$ by $\bM_2$ for computing $\wh \bA_2$.
Empirical evidences in modelling high-dimensional time series indicate that the estimation
is not sensitive to the choice of $k_0$, small values of $k_0$ such as 1 to 5 are sufficient for most applications (Lam \etal 2011, Lam and Yao 2012, Chang \etal 2015).
Since using $\bM_1$ and $\bM_2$ does not add anything fundamentally new, we
proceed with the simple version only. 
 
(ii) The proposed procedure encapsulates all the dependence across space and time into
$d$ latent factors. Those latent factors, specified objectively by sample covariances
(and autocovariances) of the data,  capture all the
linear correlations parsimoniously. The real data example in Section 6.2
below, and also those not shown in this paper, indicate that the
estimated $d$ is often small.

\end{remark}

\subsection{Aggregating via random partitioning}
\label{sec32}

The estimation for the latent variable $\xi_t(\cdot)$ depends on partitioning
$ \calS_o = \{ \bs_1, \cdots, \bs_p\}$ into two non-overlapping sets
$\calS_1$ and $\calS_2$; see (\ref{f7}). Since the estimation procedure presented
in Section \ref{sec31} puts  $\calS_1$ and $\calS_2$ on equal footing, we set
$p_1=[p/2]$ and $p_2=p-p_1$. 
By randomly dividing $ \calS_o$ into $\calS_1$ and $\calS_2$ with the sizes $p_1$
and $p_2$ respectively, the estimates for $\bxi_{t,1}$ and $\bxi_{t,2}$ are
obtained as in (\ref{f7}). We repeat this randomization $J$ times, where $J \ge 1$ is a
large integer, leading to the $J$ pairs of the estimates  $(\wh
\bxi_{t,1}^{j}, \, \wh \bxi_{t,2}^{j})$ for $j=1, \cdots, J.$
The aggregating estimator over the randomized partitions is
\begin{equation} \label{f8}
\wt  \xi_{t}(\bs_i) = {1 \over J} \sum_{j=1}^J \wh\xi_{t}^{j}(\bs_i), \qquad j=1, \cdots, p,
\end{equation}
where $\wh\xi_{t}^{j}(\bs_i)$ is a component of either $\wh \bxi_{t,1}^{j}$ or
$\wh \bxi_{t,2}^{j}$, depending on $\bs_i \in \calS_1 $ or $\calS_2$
in the $j$-th randomized partition of $\calS_o$. Similar to the Bagging method
of Breiman (1996), the choice
of $J$ is not critical. In our numerical experiments, we set $J=100$.

\begin{theorem} \label{prop2}
For $k=1, \cdots, n$ and $\ell=1, \cdots, p$,
\begin{equation} \label{p1}
E\Big(\big\{ \wt \xi_{k}(\bs_\ell) - y_{k}(\bs_\ell)\big\}^2 \Big| \{ y_t (\bs_i) \} \Big)
\le E\Big(\big\{ \wh \xi_{k}(\bs_\ell) - y_{k}(\bs_\ell)\big\}^2 \Big| \{
y_t (\bs_i) \} \Big),
\end{equation}
 and
 \begin{equation} \label{p2}
 \mathrm{E}\Big({1 \over np} \sum_{t=1}^n\sum_{j=1}^p \big\{ \wt \xi_{t}(\bs_j) - \xi_{t}(\bs_j)\}^2\Big| \{\xi_{t}(\bs_i), \, y_t (\bs_i) \} \Big)\le \mathrm{E}\Big({1 \over np} \sum_{t=1}^n\sum_{j=1}^p \big\{ \wh \xi_{t}(\bs_j) - \xi_{t}(\bs_j)\}^2\Big| \{\xi_{t}(\bs_i),\, y_t (\bs_i) \} \Big).
 \end{equation}
\end{theorem}

Theorem \ref{prop2} is in the same spirit as Breiman's inequality
for Bagging; see (4.2) in Breiman (1996).
Note that 
all the conditional expectations
in Theorem~\ref{prop2} above are taken with respect to the random partitioning
of the location set $\calS_o$ into $\calS_1 $ and $\calS_2$.
There are in total $p_0\equiv p!/(p_1!p_2!)$ different partitions, each being taken
with probability $1/p_0$. Denote by $\wh \xi_k^{(1)}(\cdot), \cdots, \wh
\xi_k^{(p_0)}(\cdot)$ the resulting $p_0$ estimates as in (\ref{f7}). Then
\begin{align} \label{c9na}
& E\Big(\big\{ \wt \xi_{k}(\bs_\ell) - y_{k}(\bs_\ell)\big\}^2 \Big| \{
y_t (\bs_i) \} \Big)\nn
\; =\; E\Big(\big\{ {1\over J}\sum_{l=1}^{J}(\wh \xi_{k}^{l}(\bs_\ell) - y_{k}(\bs_\ell))\big\}^2 \Big| \{ y_t (\bs_i) \} \Big)\nn\\
\le \; &  E\Big({1\over J}\sum_{l=1}^{J}\big\{ (\wh \xi_{k}^{l}(\bs_\ell) - y_{k}(\bs_\ell))\big\}^2 \Big| \{ y_t (\bs_i) \} \Big)\nn\\
= \;&{1 \over p_0}  \sum_{j=1}^{p_0} \big\{\wh \xi_k^{(j)} (\bs_\ell) - y_{k}(\bs_\ell)\big\}^2
= E\Big(\big\{ \wh \xi_{k}(\bs_\ell) - y_{k}(\bs_\ell)\big\}^2 \Big| \{ y_t (\bs_i) \} \Big).
\nonumber
\end{align}
This completes the proof for (\ref{p1}). Note that (\ref{p2}) can be established
in the same manner.

 \subsection{Scalable to large datasets} \label{sec33}
The estimator $\wh \bA_1$ 
in (\ref{f4}) was obtained from an eigenanalysis
which requires $O(p_1 p_2^2)$ operations. This is computational challenging when $p$ is large.
However our approach can be easily adapted to large $p$, which is in the spirit of
`divide and conquer'.

We randomly divided $\calS_0$
into $p/q$ sets $\calS_1^*, \cdots, \calS_q^*$, and each $\calS_i^*$
contains $q$ locations, where $q$ is an integer such that the eigenanalysis for
$q\times q$ matrices
can be performed comfortably with the available computing capacity.
We estimate $\xi_t(\cdot)$ at the $q$ locations in $\calS_i^*$ for each of $i=1, \cdots, p/q$
separately using the aggregation algorithm below.
\begin{quote}
(i) Randomly select $q$ locations from $\calS_0 - \calS_i^*$.\\
(ii) Combine the data on the locations in $\calS_i^*$ and the locations selected in (i). By
treating the combined data as the whole sample, calculate $\wh \xi_t(\bs)$ for $\bs \in  \calS_i^*$
as in (\ref{f7}).\\
(iii) Repeat (i) and (ii) above $J$ times, aggregate the estimates as in (\ref{f8}).
\end{quote}

Alternatively, we can randomly choose $2q$ locations from $\calS_0$ to perform the estimation
(\ref{f7}). Repeating the estimation a large number (say, greater than $Jp/(2q)$)
of times, we then aggregate the estimates
at each location as in (\ref{f8}). This is a computationally more efficient approach with the
drawback that the number of the estimates obtained at each location  is not directly under
control.


\subsection{Regression estimation} \label{sec34}

In the presence of observable covariant $\bz_t(\cdot)$ in (\ref{b1}), the
regression coefficient vector $\bbeta(\cdot)$
can be estimated by the least squares method. To this end, let
\begin{equation} \label{c9}
\by(\bs_i) = (y_1(\bs_i), \cdots, y_n(\bs_i))', \qquad
\bZ(\bs_i)= (\bz_1(\bs_i), \cdots, \bz_n(\bs_i) )'.
\end{equation}
It follows from (\ref{b1}) that
\[
\by(\bs_i) = \bZ(\bs_i) \bbeta(\bs_i) + \be(\bs_i),
\]
where $\be(\bs_i) = (\xi_1(\bs_i)+ \ve_1(\bs_i), \cdots, \xi_n(\bs_i)+ \ve_n(\bs_i))'$. Thus
the least squares estimator for $\bbeta(\bs_i)$ is defined as
\begin{equation} \label{c10}
\wh \bbeta(\bs_i) = \{ \bZ(\bs_i)' \bZ(\bs_i)\}^{-1} \bZ(\bs_i)' \by(\bs_i), \qquad i=1, \cdots, p.
\end{equation}
Then by replacing the original data $y_t(\bs_i)$ by the regression
residuals $ y_t(\bs_i) - \bz_t(\bs_i)' \wh \bbeta(\bs_i)$, we proceed to estimate the finite dimensional
structure of $\xi_t(\cdot)$ as described in Section \ref{sec31} above.

However in the presence of the endogeneity in the sense $\cov(\bz_t(\bs), \,
\xi_t(\bs)) \ne 0$, the
regression estimator $\wh \bbeta(\bs_i)$ in (\ref{c10})  is
practically an estimator for
$$\bbeta(\bs_i)^\star \equiv \bbeta(\bs_i) + \var(\bz_t(\bs_i))^{-1}
\cov(\bz_t(\bs_i),\, \xi_t(\bs_i))$$
instead,
as (\ref{b1}) can be written as
$
y_t(\bs) =  \bz_t (\bs)'  \bbeta(\bs)^\star + \xi_t(\bs)^\star  + \ve_t(\bs)
$,
where  $$\xi_t(\bs)^\star  = \xi_t(\bs) -
\bz_t (\bs)' \var(\bz_t(\bs_i))^{-1} \cov(\bz_t(\bs_i), \xi_t(\bs_i)).$$
It is easy to see that
$
\cov( \bz_t (\bs), \; \xi_t(\bs)^\star  )
=0$. Hence $\wh \bbeta(\bs_i)$ is a consistent estimator for
$\bbeta(\bs_i)^\star$. Furthermore, the estimation
based on the residuals described above is still valid though the finite
dimensional structure (\ref{b4})  is
now imposed upon the latent process $\xi_t(\bs)^\star$ instead.

\section{Kriging}
\label{sec4}

First we state a general lemma on linear prediction  which shows explicitly
the terms required in order to carry out kriging for spatio-temporal
process $y_t(\bs)$.

\askip

\begin{lemma} For any random vectors  $\bzeta$ and $\bfeta$ with
$E( \|\bzeta\|^2 + \|\bfeta\|^2) < \infty$, the best linear predictor for $\bzeta$ based on
$\bfeta$ is defined as  $\wh \bzeta = \balpha_0 + \bB_0 \bfeta$, where
\[
(\balpha_0, \bB_0) = \arg \inf_{\balpha, \bB}
E\big\{ \| \bzeta-\balpha - \bB \bfeta\|^2 \big\}.
\]
In fact,
\[
\bB_0 = \cov(\bzeta, \bfeta)\{ \var(\bfeta) \}^{-1} , \qquad \balpha_0 =
E \bzeta - \bB_0 E\bfeta.
\]
Furthermore,
\begin{equation} \label{d0}
E\{( \wh \bzeta - \bzeta) ( \wh \bzeta - \bzeta)' \} = \var(\bzeta) - \cov(\bzeta,
\bfeta)\{ \var(\bfeta) \}^{-1}
\cov(\bfeta, \bzeta).
\end{equation}
\end{lemma}
\askip

With the above lemma, we can predict any value $y_t(\bs)$.
With two scenarios considered below, we illustrate how to calculate
inverses of large covariance matrices by taking advantages from the
finite dimensional structure (\ref{b4}): all matrices to be inverted are
of the sizes $d\times d$ 
only, regardless the size of $p$.
Technically we repeatedly use the following formulas for the inverses of
partitioned matrices.

\askip

\begin{lemma}
For an invertible block-partitioned matrix $\bH= \Big(
\begin{array}{cc}
\bH_{11} & \bH_{12} \\
\bH_{21} & \bH_{22}
\end{array}
\Big)$, it holds that
\begin{equation} \label{d1}
\bH^{-1} =  \Big(
\begin{array}{ll}
\bH_{11}^{-1} + \bH_{11}^{-1}\bH_{12}(\bH_{22} - \bH_{21} \bH_{11}^{-1}\bH_{12})^{-1}
\bH_{21} \bH_{11}^{-1} & - \bH_{11}^{-1}\bH_{12} (\bH_{22} - \bH_{21}
\bH_{11}^{-1}\bH_{12})^{-1}\\
-(\bH_{22} - \bH_{21}
\bH_{11}^{-1}\bH_{12})^{-1} \bH_{21} \bH_{11}^{-1} &
(\bH_{22} - \bH_{21}
\bH_{11}^{-1}\bH_{12})^{-1}
\end{array}
\Big)
\end{equation}
provided $\bH_{11}^{-1}$ exists. Furthermore,
\begin{equation} \label{d2}
(\bH_{22} - \bH_{21}
\bH_{11}^{-1}\bH_{12})^{-1} =
\bH_{22}^{-1} + \bH_{22}^{-1}\bH_{21}(\bH_{11} - \bH_{12} \bH_{22}^{-1}\bH_{21})^{-1}
\bH_{12} \bH_{22}^{-1}
\end{equation}
provided both $\bH_{11}^{-1}$ and $\bH_{22}^{-1}$ exist.
\end{lemma}
\askip

Formula (\ref{d1}) can be proved by checking $\bH^{-1} \bH = \bI$ directly, while
(\ref{d2}) follows from (\ref{d1}) by comparing the (1,1) and (2,2) blocks on the
RHS of (\ref{d1}).

\subsection{Kriging over space}
\label{sec41}

The goal is to predict the unobserved value $y_{t}(\bs_0)$ for some $\bs_0 \in \calS$,
$1\le t \le n$, and $\bs_0 \ne \bs_j $ for $1\le j \le p$, based on the observations
$\by_t \equiv  (\by_{t,1}', \by_{t,2}')'$ only, where $\by_{t,1}, \by_{t,2}$ are
defined as in (\ref{f1}).
We introduce two predictors below.
We always use the notation
$K_h(\cdot) = h^{-1} K(\cdot/h)$, where $K(\cdot)$ denotes a kernel function, $h>0$ is a
bandwidth, and $K$ and $h$ may be different at different places.

To simplify the notation, we assume $\bbeta(\bs) \equiv 0$ in (\ref{b1}). As
indicated in Section
\ref{sec34}, this effectively implies to replace the observations $y_t(\bs_j)$ by
the regression residuals. For kriging, we also need to estimate $\bbeta(\bs_0)$ based on
$\wh \bbeta(\bs_j)$, $j=1, \cdots, p$, given in (\ref{c10}). It can be achieved by, for example,
using the kernel smoothing:
\begin{equation} \label{d6}
\wh \bbeta(\bs_0) = \sum_{j=1}^p \wh \bbeta(\bs_j) K_h(\bs_j - \bs_0) \Big/
\sum_{j=1}^p   K_h(\bs_j - \bs_0),
\end{equation}
where $K(\cdot)$ is a density function defined on $\RR^2$, $h>0$ is a bandwidth.
  Furthermore, a local linear smoothing can be applied to
improve the accuracy of the estimation; see, e.g. Chapter 3 of Fan and Gijbels (1996).
By the standard argument it can be shown (see the supplementary document) that
\[
|\wh \bbeta(\bs_0) - \bbeta(\bs_0)| = O_p(h^2 + n^{-1/2}),
\]
provided that the conditions in Theorem \ref{th3} in Section \ref{sec52} below hold.
Note that if $\bbeta(\bs_j)$, $j=1, \cdots, p$,  were all known, the above error rate
reduces to $O_p(h^2)$, as $\bbeta(\cdot)$ is deterministic
and continuous. See Condition
4 in Section \ref{sec52} below. The term of order $n^{-1/2}$ reflects the
errors in estimation for
$\bbeta(\bs_j)$.
In the rest of Section \ref{sec4}, we adhere with the
assumption $\bbeta(\bs) \equiv 0$.


\askip

It follows from Lemma 1 that the best linear predictor for $y_{t}(\bs_0)$ based on
$\by_t $ is
\begin{equation} \label{d5}
\wh y_t(\bs_0) = \cov(y_t(\bs_0), \by_t)  \var(\by_t)^{-1} \by_t.
\end{equation}
It follows from (\ref{d0}) that
\begin{align} \nonumber
& E[\{  \wh y_t(\bs_0) - y_t(\bs_0) \}^2] = \var\{y_t(\bs_0)\}
- \cov(y_t(\bs_0), \by_t) \var(\by_t)^{-1} \cov(\by_t, y_t(\bs_0)) \\
= \; & \sigma(\bs_0)^2 + \var\{ \xi_t(\bs_0) \}
- \cov(\xi_t(\bs_0), \bxi_t) \{  \var(\bxi_t) + \bD \}^{-1} \cov(\bxi_t, \xi_t(\bs_0)),
\label{f11}
\end{align}
where  $\bD =\var(\bve_t)$ is a diagonal matrix,
$\bve_t = (\bve_{t,1}', \bve_{t,2}')'$ and $\bxi_t = (\bxi_{t,1}', \bxi_{t,2}')'$. See (\ref{f1}).

To apply predictor $\wh y_t(\bs_0)$ in (\ref{d5}) in practice, we need to estimate both
$\cov(y_t(\bs_0), \by_t)$ and $\var(\by_t)$. Since
$
\cov(y_t(\bs_0), \by_t) = \cov(\xi_t(\bs_0), \by_t)
$,
it can be estimated by
\[
c(\bs_0) = {1 \over n} \sum_{k=1}^n (\wh \xi_k(\bs_0) - \bar \xi(\bs_0) )
(\by_k - \bar \by),
\]
where $\wh \xi_t(\bs_0)$ is a kernel estimator for $\xi_t(\bs_0)$ defined as
\begin{equation} \label{f10n}
\wh \xi_t(\bs_0) = \sum_{j=1}^p \wh \xi_t(\bs_j) K_h(\bs_j - \bs_0)
\Big/ \sum_{j=1}^p K_h(\bs_j - \bs_0)
\end{equation}
with $\wh \xi_t(\bs_1), \cdots
\wh \xi_t(\bs_p)$ defined in (\ref{f7}) (see also (\ref{d6}) above), and
$\bar \xi(\bs_0) =  n^{-1} \sum_t \wh \xi_t(\bs_0)$.
Thus a realistic predictor for $y_t(\bs_0)$ is
\begin{equation} \label{f10}
\wh y_t^r(\bs_0)=c(\bs_0) \wh \bSigma_y^{-1} \by_t,
\end{equation}
where $\wh \bSigma_y = n^{-1} \sum_{k=1}^n (\by_k -\bar \by)(\by_k -\bar \by)'$ is the sample
variance of $\by_t$. Nevertheless it turns out that
\begin{equation} \label{f10new}
\wh y_t^r(\bs_0)=\wh \xi_t(\bs_0) .
\end{equation}
To show this,
let $w_j=K_h(\bs_j - \bs_0)
\big/ \sum_{j=1}^p K_h(\bs_j - \bs_0)$. It follows from (\ref{f7}) that
\beqn \wh y_t^r(\bs_0)&=&(w_1, \cdots, w_p)\Big[{1\over
n}\sum_{k=1}^{n}(\wh\bxi_k-\bar{\bxi})(\by_k-\bar\by)'\Big]\wh\Sigma_y^{-1}\by_t\nn\\
&=&(w_1, \cdots, w_p)\left( \begin{array}{cc} \wh \bA_1 \wh\bA'_1 & \bf0
\\ \bf0 &\wh \bA_2 \wh\bA'_2 \end{array} \right)\Big[{1\over
n}\sum_{k=1}^{n}(\by_k-\bar{\by})(\by_k-\bar\by)'\Big]\wh\Sigma_y^{-1}\by_t\nn\\
&=&(w_1, \cdots, w_p)\left( \begin{array}{cc} \wh \bA_1 \wh\bA'_1 & \bf0 \\ \bf0 &\wh \bA_2 \wh\bA'_2 \end{array} \right)\by_t\nn
=(w_1, \cdots, w_p)\left(\begin{array}{c}\wh \bxi_{t,1}\\ \wh \bxi_{t,2} \end{array}\right)=\wh \xi_t(\bs_0).\nn\eeqn

It is worth pointing out that expression (\ref{f10}) involves inverting  $p\times p$ matrix
$\wh \bSigma_y$, which is difficult when $p$ is large, while (\ref{f10new}) paves the way
for computing the predictor $\wh y^r_t(\bs_0)$ without the need to
compute $\wh \bSigma_y^{-1}$ directly.

By Theorem \ref{prop2}, a better predictor than $\wh y_t^r(\bs_0)$ in (\ref{f10new}) is
\begin{equation} \label{f12}
\wt y_t^r(\bs_0) \equiv \wt \xi_t(\bs_0)
 = \sum_{j=1}^p \wt \xi_t(\bs_j) K_h(\bs_j - \bs_0)
\Big/ \sum_{j=1}^p K_h(\bs_j - \bs_0),
\end{equation}
where $\wt \xi_t(\bs_j)$ is defined in (\ref{f8}).

Both $\wh y_t^r(\bs_0)$ and $\wt y_t^r(\bs_0)$ are the
approximate linear estimators for the
$\xi_t(\bs_0)$ based on $\xi_t(\bs_1), \cdots, \xi_t(\bs_p)$.
Note that $y_t(\bs_0) = \xi_t(\bs_0) + \ve_t(\bs_0)$,
and the nugget effect term $\ve_t(\bs_0)$ is unpredictable. The best (unrealistic)
predictor for $y_t(\bs_0)$ is $\xi_t(\bs_0)$. It is indeed recommended to
predict $\xi_t(\bs_0)$ instead of $y_t(\bs_0)$ directly. See also pp.136-137 of
Cressie and Wikle (2011).

\begin{remark} \label{remark3}
(i) The realistic kriging estimators $\wh y_t^r(\bs_0)$ and $\wt
y_t^r(\bs_0)$ actually make the full use of all the available data, in
spite that they were induced from (\ref{d5}).  Note that the ideal
(and unrealistic) preditor for $y_t(\bs_0)$ is $\sum_{1\le j \le d} a_j(\bs_0)
x_{tj}$, and $\wh x_{tj}$ and $\wt x_{tj}$ are the estimators 
for $x_{tj}$ based on all the available data from time 1 to $n$.
It follows from (\ref{f10n}) -- (\ref{f10new})
that $\wh y_t^r(\bs_0)$ is a realistic optimal predictor for $y_t(\bs_0)$ based on
$\{\, \wh x_{tj},\,  j=1, \cdots \wh d \,\}$.

(ii) When the number of observations in the vicinity of $\bs_0$ is small, the kernel
based predictor (\ref{f10n}) may perform poorly. One alternative is to impose a
parametric spatial covariance function and to perform the kriging based on the parametric
model (Sections 4.1.1 and 6.1 of Cressie and Wikle, 2011). How to identify an appropriate 
parametric model using the nonparametric analysis presented in this paper deserves 
a separate study.
  
\end{remark}

\subsection{Kriging in time}
\label{sec42}

\subsubsection{Prediction methods}
The goal now is to predict the future values $y_{n+j}(\bs_1),
\cdots, y_{n+j}(\bs_p) $,  for some $j\ge 1$, based on $\by_n, \cdots, \by_{n-j_0}$,
where $0\le j_0 < n$ is a prescribed integer. When $j_0=n-1$, we use all the
available data to predict the future values.
Since  $\ve_{t+j}(\cdot)$ is unpredictable,
a more effective approach is to predict  $\bx_{n+j}= (x_{n+j,1}, \cdots,
x_{n+j, d})'$ based on $\bx_n,
\cdots, \bx_{n-j_0}$, as the ideal predictor for $y_{n+j}(\bs_i)$ is
$\xi_{n+j}(\bs_i)$;
see (\ref{c2}).

Since our procedure to recover the latent process $\bx_t$ requires to split $\by_t$ into
two subvectors $\by_{t,1}, \; \by_{t,2}$, leading to two different configurations
$ \bx_t$ and $ \bx_t^\star$ in (\ref{f2}), we will apply the
prediction procedure in Section \ref{sec422} below to each of $ \bx_t$ and $ \bx_t^\star$.
Then the predictors for $\by_{n+j, 1}$ and $\by_{n+j,2}$ are defined as
\begin{equation}\label{f12nn}
 \by_{n,1}(j) =  \bA_1  \bx_n(j), \qquad  \by_{n,2}(j) =  \bA_2  \bx_n^\star(j),
\end{equation}
where $\bx_n(j)$ is the predictor for $\bx_{n+j}$,
and $ \bx_n^\star(j)$ is the predictor for $ \bx_{n+j}^\star$.
In practice, $\bA_i, \bx_t, \bx_t^\star$ are replaced by their estimators defined in
(\ref{f4}) and (\ref{f6}).

The predictors defined above depend on a single partition $\calS_o= \calS_1 \cup \calS_2$.
By repeating random partition of $\calS_o$ $J$ times, we may obtain the aggregated predicted
values for $y_{n+j}(\bs_i)$ in the same manner as in (\ref{f8}).

Since $\xi_t(\bs_1), \cdots, \xi_t(\bs_p)$ are correlated with each other, we should not
model $\xi_t$ at each location separately. Instead modeling the factor process $\bx_t$
catches the temporal dynamics much more parsimoniously.

An alternative approach, not pursued here, would be to build a dynamic model for
$\bx_t$, leading to the model-bases forecasts.
For example, Cressie, Shi and Kang (2010) adopt the Gaussian AR(1)
specification for the latent process and facilitated the forecasting
by a Kalman filter.

\subsubsection{Predicting $\bx_{n+j}$ and $\bx_{n+j}^\star$}
\label{sec422}

We only state the method for predicting $\bx_{n+j}$. It can be applied to  predicting
$\bx_{n+j}^\star$ exactly in the same manner.

Let $\bX' = (\bx_n', \cdots, \bx_{n-j_0}')$,
\begin{equation} \label{d11}
\bW_k \equiv
\var\left(
\begin{array}{l}
\bx_t\\
\bx_{t-1}\\
\vdots\\
\bx_{t-k}
\end{array}
\right)
=
\left(
\begin{array}{llll}
\bSigma_x(0) & \bSigma_x(1) & \cdots & \bSigma_x(k) \\
\bSigma_x(1)' & \bSigma_x(0) & \cdots & \bSigma_x(k-1) \\
& \cdots& \cdots& \\
\bSigma_x(k)' & \bSigma_x(k-1)'& \cdots &\bSigma_x(0)
\end{array}
\right), \quad k\ge 0,
\end{equation}
\[
\bR_{j_0}\equiv \left( \bSigma_x(j), \bSigma_x(j+1).
 \cdots,
\bSigma_x(j+j_0)
\right),
\]
where $\bSigma_x(k) = \cov(\bx_{t+k}, \bx_t)$.
By Lemma 1, the best linear predictor for $\bx_{n+j}$ is
\[
\bx_{n}(j) = \bR_{j_0}
\bW_{j_0}^{-1} \bX.
\]
The key is to be able to calculate the inverse of $(j_0+1)d \times (j_0+1)d$ matrix
$\bW_{j_0}$. This can be done by calculating $\bW_0^{-1}, \bW_1^{-1}, \cdots $
recursively based on
\begin{equation} \label{d12}
\bW_{k+1}^{-1}
= \Big(
\begin{array}{ll}
\bW_k^{-1} + \bW_k^{-1} \bU_k \bV_k \bU_k' \bW_k^{-1} & - \bW_k^{-1}\bU_k \bV_k\\
- \bV_k \bU_k' \bW_k^{-1} & \bV_k
\end{array}
\Big),
\end{equation}
where
$$ \bU_k' = ( \bSigma_x(k+1)', \cdots, \bSigma_x(1)'), \qquad
\bV_k = (\bSigma_x(0) -\bU_k'\bW_k^{-1}\bU_k)^{-1}.$$
See (\ref{d1}). Note only $d\times d$ inverse matrices are involved in this
recursion.

In practice we replace $\bSigma_x(k)$ in $\bR_{j_0}$ and $\bW_{j_0}$ by
$
\wh \bSigma_x(k) =  \wh \bA_1'\wh \bSigma_{y,1}(k) \wh \bA_1,
$
and replace $\bX$ by
\[
\wh \bX = (\by_{t,1}'\wh\bA_1, \cdots, \by_{t-k,1}'\wh\bA_1)',
\]
where
\[
\wh \bSigma_{y,1}(k)= {1 \over n} \sum_{t=1}^{n-k}(\by_{t+k, 1} - \bar \by_1)
(\by_{t, 1} - \bar \by_1)' , \qquad \bar \by_1 = {1 \over n} \sum_{t=1}^n \by_{t,1}.
\]
The resulting predictor for $\bx_{n+j}$ is denoted by
$\wh \bx_{n}(j)$.

We may define $\wh \bx_{n}^\star(j)$ in the same manner as $\wh \bx_{n}(j)$
with $(\by_{t,1}, \wh \bA_1)$ replaced by $(\by_{t,2}, \wh \bA_2)$.
 Consequently the practical feasible predictor for $\by_{n+j}$ is
defined in two similar formulas
\begin{equation} \label{f12n}
\wh y_{n,1}(j) = \wh \bA_1 \wh \bx_{n}(j), \qquad
\wh y_{n,2}(j) = \wh \bA_2 \wh \bx_{n}^\star(j),
\end{equation}
see (\ref{f12nn}).

\subsection{Handling missing values}
\label{sec43}

It is not uncommon that a large data set contains some missing values.
We assume that the number of missing values is small in the sense that
the number of the available observations at each given time $t$ is of the order
$p$, and the number of the available observations at each location $s_i$ is of
the order $n$. 	
We outline below how to apply the proposed method when some observations
are missing.

First for $\bSigma\equiv ( \sigma_{ij})$ defined in (\ref{f5n}), we may estimate
each $\sigma_{ij}$ separately using all the available pairs $(y_{t,i}^1, y_{t,j}^2)$
with $1\le t \le n$, where $y_{t,i}^{\ell}$ denotes the $i$-th element of $\by_{t, \ell}$,
$\ell =1, 2$. With the estimated $\wh \bSigma$, we may derive the estimators
$\wh \bA_1, \, \wh \bA_2$ as in (\ref{f4}). 		

For the simplicity in notation, suppose that $y_1(\bs_1)$ is missing.
Let $\by_1^a$ denote all the available observations at time $t=1$. By Lemma 1,
the kriging predictor for $y_1(\bs_1)$ is
\begin{equation} \label{xx1}
\wh y_1(\bs_1) = \cov(y_1(\bs_1), \by_1^a) \{\var(\by_1^a) \}^{-1} \by_1^a.
\end{equation}
We may estimate $\cov(y_1(\bs_1), \by_1^a)$ and $\var(\by_1^a)$ in the same
manner as that for estimating $\bSigma$ described above. Replacing all the missing
values with their kriging estimates, we may proceed the estimation for $
\wh \xi_t(\bs_j)$ and $\wt \xi_t(\bs_j)$ as in Sections \ref{sec31} and \ref{sec32}.


\section{Asymptotic properties}
\label{sec5}

In this section, we investigate the asymptotic properties of  the
proposed methods.
For any matrix $\bM$,
let $||\bM||_{\min}=\sqrt{\lambda_{\min}(\bM\bM')}$ and $||\bM||=\sqrt{\lambda_{\max}(\bM\bM')}$, where
$\lambda_{\min}$ and  $\lambda_{\max}$ denote, respectively, the minimum and the maximum
eigenvalue.
When $\bM$ is a vector, $||\bM||$ reduces to its Euclidean norm.

\subsection{On latent finite-dimensional structures}
\label{sec51}

We state in this subsection some asymptotic results
 on the estimation of the
factor loading spaces $\calM(\bA_1)$ and $\calM(\bA_2)$. They paves the way to
establish the properties for the kriging estimation presented in Section
\ref{sec52} below.
Proposition \ref{them52} below is  similar to those  in Lam and
Yao (2012),  and Chang \etal (2015) but with the extra features due
to the graph Laplician incorporated
in order to pertain the continuity over space.
Nevertheless its proof is similar and, therefore, is omitted.

For any two $k\times d$ orthogonal matrices $\bB_1$ and $ \bB_2$ with
$\bB_1' \bB_1=\bB_2' \bB_2=\bI_d$,
we measure the distance between the two linear spaces $\calM(\bB_1)$ and $\calM(\bB_2)$ by
\begin{equation} \label{5.1}
D(\mathcal{M}(\bB_1), \mathcal{M}(\bB_2))=\sqrt{1-{1\over
d}\mathrm{tr}(\bB_1\bB_1' \bB_2 \bB_2')}.
\end{equation}
It can be shown that $D(\mathcal{M}(\bB_1), \mathcal{M}(\bB_2)) \in [0, 1]$, being 0
if and only if ${\mathcal{M}}(\bB_1) ={\mathcal{M}}(\bB_2)$,
and 1 if and only if ${\mathcal{M}}(\bB_1)$ and ${\mathcal{M}}(\bB_2)$ are orthogonal.
We introduce some regularity conditions first. Put
\[
\by_t = ( y_t(\bs_1), \cdots, y_t(\bs_p))', \qquad
\bZ_t = ( \bz_t(\bs_1), \cdots, \bz_t(\bs_p)).
\]

\begin{quote}
\noindent {\bf Condition 1}. $\{ (\by_t, \bZ_t), \, t=0, \pm 1, \pm 2, \cdots\} $ is a strictly
stationary and $\alpha$-mixing process with  $\max_{1\leq i\leq
p}[\mathrm{E}|y_t(\bs_i)|^{\gamma}+\mathrm{E}||\bz_t(\bs_i)||^{\gamma}]<\infty$ for
some $\gamma> \max\{\beta, 4\}, \, \beta>2$ and the $\alpha$-mixing coefficients $\alpha_m$
satisfying the condition
\beqn \label{5.2}\alpha_m=O(m^{-\theta}) \quad \hbox{for some}\quad \theta> {\gamma \beta/(\gamma-\beta)}.\eeqn
Further, $\min_{1\leq i\leq p}\lambda_{\min}(\mathrm{Var}(\bz_t(\bs_i)))>c_0$ for some positive constant $c_0$.

\noindent {\bf Condition 2.} Let $\bSigma_x = \cov(\bx_t, \bx_t^\star)$, where
$\bx_t$ and $\bx_t^\star$ are defined in (\ref{f2}).
There exists a constant $\delta\in [0, 1]$ for which
$||\bSigma_x||_{\min}\asymp ||\bSigma_x||\asymp p^{1-\delta}$.

\end{quote}

\askip

Constant $\delta$ in Condition 2 reflects the strength of factors.
Intuitively a strong factor is linked
with most components of $\by_{t,1}$ and $\by_{t,2}$, implying that the corresponding
coefficients in $\bA_1$ or $\bA_2$ are non-zero.
Therefore it is relatively easy to recover those strong factors from the observations.
Unfortunately the mathematical definition of the factor strength
is tangled with the standardization
condition $\bA_1' \bA_1 =\bA_2' \bA_2= \bI_d$. See Remark 1(i) of Lam and Yao (2012), and
Lemma 1 of Lam \etal (2011).
 To simplify the presentation,
Condition 2 assumes that
all the factors in (\ref{f2}) are of the same strength which is measured by a
constant $\delta \in [0, 1]$:
$\delta=0$ indicates that the strength of the factors is at its strongest, and
$\delta=1$ corresponds to the weakest factors.


\begin{proposition} \label{them52} Let
Conditions 1 and 2 hold, and $p^{\delta}n^{-1/2}+p
n^{-\beta/2}+p^{2\delta-2}\tau \|\bL\| \to 0$ as
$n\to \infty$. Then
\begin{itemize}
\item [(i)] $|\hat{\lambda}_i-\lambda_i|=O_p(p^{2-\delta}n^{-1/2}+\tau \|\bL\|)$ for $1 \le i \le  d,$

\item [(ii)] $|\hat{\lambda}_i|=O_p(p^2n^{-1}+\tau \|\bL\|)$ for $d<i\leq p$,  and

\item [(iii)] $D(\mathcal{M}(\wh\bA_i), \mathcal{M}(\bA_i)) = O_p(p^\delta n^{-1/2}+
p^{2\delta-2}\tau \|\bL\|)$
($ i=1, 2$), provided that $d$ is known.
 \end{itemize}
 \end{proposition}

%

\begin{remark} \label{remark2}
(i)
  Proposition \ref{them52} indicates that stronger factors result in
a better estimation for the factor loading spaces, and, consequently,
 a better recovery of the factor process.
This is due to the fact that
$\lambda_d - \lambda_{d+1}$ increases as $\delta$ decreases, where $\lambda_i$ denotes the $i$-th
largest eigenvalue of $\bSigma \bSigma'$, and $\bSigma$ is defined in (\ref{f5n}).
Especially with the strongest factors
(i.e. $\delta=0$), $D(\mathcal{M}(\wh\bA_i), \mathcal{M}(\bA_i))$
attains the standard error rate $n^{-1/2}+p^{-2}\tau\|\bL\|$.
This phenomenon is coined as `blessing of dimensionality' as in Lam and Yao (2012).

(ii) Proposition \ref{them52}(iii) can be made adaptive to unknown $d$; see Remark 5 of
Bathia \etal (2010). See also Theorem 2.4 of Chang \etal (2015) on how to make $\wh d $
defined in (\ref{c5}) be a consistent estimator for $d$.

(iii) The condition $p^{2\delta-2}\tau \|\bL\|\rightarrow 0$
in Proposition \ref{them52}
 controls the perturbation between $\wh\bA$ and $\bA$, which is implied by
either $p^{2\delta-1}\tau \to 0$ (as $\|\bL\|\le p$) or $\|\bL\|\le C$
and $p^{2\delta-2}\tau \to 0$. 
By
the perturbation theory (Theorem 8.1.10 of Golub and Van Loan, 1996),
the bound $||\wh\bA-\bA||$ depends on $||\wh\bSigma \wh\bSigma'+\tau
\bL-\bSigma\bSigma'||$, which is bounded from above by
$ ||\wh\bSigma \wh\bSigma'-\bSigma\bSigma'||+\tau||\bL||.$  This leads to
the upper bound of (iii)  in Proposition \ref{them52}.

\end{remark}

\subsection{On kriging}
\label{sec52}

We now consider the asymptotic properties for the kriging
methods proposed in Section \ref{sec4}. To simplify the presentation,
we always assume that $d$ is known. We introduce some regularity conditions first.

\begin{quote}
\noindent {\bf Condition 3}. The kernel $K(\cdot)$ is a symmetric density
function on $\RR^2$ with a bounded support.
\end{quote}

\begin{quote}
\noindent {\bf Condition 4}. In (\ref{c2}) $ \bbeta(\cdot)$ and $
a_j(\cdot)/||\ba(\bs_0)||, \, j=1, \cdots, d$,
are  twice continuously differentiable and bounded functions on $\calS$, where $\ba(\bs_0)=(a_1(\bs_0), \cdots, a_d(\bs_0)).$
\end{quote}

\begin{quote}
\noindent {\bf Condition 5}. There exists a positive and continuously differentiable sampling intensity  $f(s)$ on $\calS$ such that as $p\rightarrow \infty,$
\beqn {1\over p}\sum_{\bs \in \calS} I(\bs \in A) =\int_{A}
f(\bs) \, d\bs(1+o(1))\nn\eeqn
holds for any measurable set $A\subset \calS$.
\end{quote}

Theorem \ref{th3} below presents the asymptotic properties of the two
spatial kriging methods in (\ref{f10new}) and (\ref{f12}).
Since $$E[\{  \wh y_t^r(\bs_0) - y_t(\bs_0) \}^2]
=E[\{  \wh y_t^r(\bs_0) -\xi_t(\bs_0)\}^2]
+ \var(\ve_t(\bs_0)),$$
it is more relevant
to measure the difference between a predictor and $\xi_t(\bs_0)$ directly.

\begin{theorem} \label{th3} Let bandwidth $h\rightarrow 0, \,  ph\rightarrow\infty$ and   $p^{\delta}n^{-1/2}+p n^{-\beta/2}+p^{2\delta-2}\tau \|\bL\| \to 0$ as
$n\to \infty$.
It holds under Conditions 1--5  that
 $$\max\{|\wh y^r_t(\bs_0)-\xi_t(\bs_0)|, \, |\wt
y^r_t(\bs_0)-\xi_t(\bs_0)|\}=O_p\{h^2+p^\delta (nh)^{-1/2}+(ph)^{-1/2}+p^{2\delta-2}h^{-1/2}\tau\|\bL\|\}.$$

\end{theorem}

\askip


Theorem \ref{th4} below considers the convergence rates for the kriging
predictions in time. Recall $\wh \by_{n,1}(j), \;
\wh \by_{n,2}(j), \; \wh \bx_{n}(j)$ and $\wh \bx^\star_{n}(j)$ as defined
in (\ref{f12n}).

\begin{theorem} \label{th4}
Let  Conditions 1 and 2 hold. As  $n, p \to \infty $ and
$ p^{\delta/2}(p^{\delta}n^{-1/2}+p^{2\delta-2}\tau \|\bL\|) \to 0$,
\begin{itemize}
    \item[] (a)  $p^{-{1\over 2}}||\wh \bx_{n}(j)-\bx_{n}(j)||=O_p(p^{\delta}n^{-1/2}+p^{2\delta-2}\tau\|\bL\|+p^{-{1\over 2}}),$ \\
    $p^{-{1\over 2}}||\wh \bx_{n}^\star(j)-\bx^\star_{n}(j)||=O_p(p^{\delta}n^{-1/2}+p^{2\delta-2}\tau\|\bL\|+p^{-{1\over 2}})$, and
    \item[] (b) $
    p^{-{1\over 2}} ||\wh \by_{n, i}(j)- \by_{n,i}(j)||=O_p(p^{\delta}n^{-1/2}+p^{2\delta-2}\tau\|\bL\|+p^{-{1\over 2}})$ for $i=1,2$.
\end{itemize}
\end{theorem}

Theorems \ref{th3} and \ref{th4} indicate that result in
better predictions. See also Remark \ref{remark2}(i) above.

\section{Numerical properties}\label{sec6}

We illustrate the finite sample properties of the proposed methods via
both simulated and real data.

\subsection{Simulation}

For simplicity,
we let $\bs_1, \cdots, \bs_p$ be drawn randomly from the uniform distribution on $[-1, 1]^2$
and $y_t(\bs_i)$ be generated from (\ref{c2}) in which $d=3$,
$\ve_t(\bs)$ are independent and standard normal,
and
\[
a_1(\bs) = s_1/2, \qquad a_2(\bs) = s_2/2, \qquad a_3(\bs) = (s_1^2+ s_2^2)/2,
\]
\[
x_{t1} = -0.8 x_{t-1,1} + e_{t1}, \qquad
x_{t2} = e_{t2} - 0.5 e_{t-1,2}, \qquad
x_{t3} = -0.6 x_{t-1,3} + e_{t3} + 0.3e_{t-1,3}.
\]
In the above expressions, $e_{ti}$ are independent and standard normal.
The signal-noise-ratio, which is defined as
$$
\frac{\int_{\bs\in[-1,1]^2} \sqrt{\rm{VAR}(\xi_t(\bs))} d\bs }
{\int_{\bs\in[-1,1]^2} \sqrt{\rm{VAR}(\varepsilon_t(\bs))} d\bs},
$$
is about 0.72.

With $n=80, 160$ or $320$, and $p=50, 100$, or $200$,
we draw 100 samples from each setting.
With each sample, we calculate $\wh d$ as in (\ref{c5}),
and the factor loadings $\wh \bA_1$ and $ \wh \bA_2$ as in (\ref{f4}).
For the latter, we choose the tuning parameter $\tau$ over 101 grid points
between 0 and 10 by a five-fold cross-validation:  we divide $\bs_1, \cdots, \bs_n$
into 5 groups of the same size. Each time we use the data at the locations in four groups
for estimation, and predict the values at the locations in the other group
by spatial kriging (\ref{f10new}). We use Gaussian kernel
in (\ref{f10n}) with bandwidth $h$
selected by leave-one-out cross validation method.

As the estimated value $\wh d$ may not always be equal to $d$,
and $\bA_1, \, \bA_2$ are not half-orthogonal matrices in the model specified above,
we extend the distance measure for two linear spaces (\ref{5.1}) as follows:
\[
D(\calM(\wh\bA_i ), \calM(\bA_i)) = \Big( 1 - {1 \over \max( d, \wh d) } \tr\{\wh \bA_i \wh \bA_i'
\bA_i (\bA_i'\bA_i)^{-1} \bA_i' \} \Big)^{1/2}.
\]
It can be shown that $D(\calM(\wh\bA_i ), \calM(\bA_i)) \in [0, 1]$,
being 0 if and only if $\calM(\wh\bA_i ) = \calM(\bA_i)$,
and 1 if and only if $\calM(\wh\bA_i)$ and $ \calM(\bA_i)$ are orthogonal.
It reduces to (\ref{5.1}) when $\wh d = d$ and $\bA_i'\bA_i = \bI_d$.

\begin{figure}
\begin{center}
    \psfig{angle=-90,figure=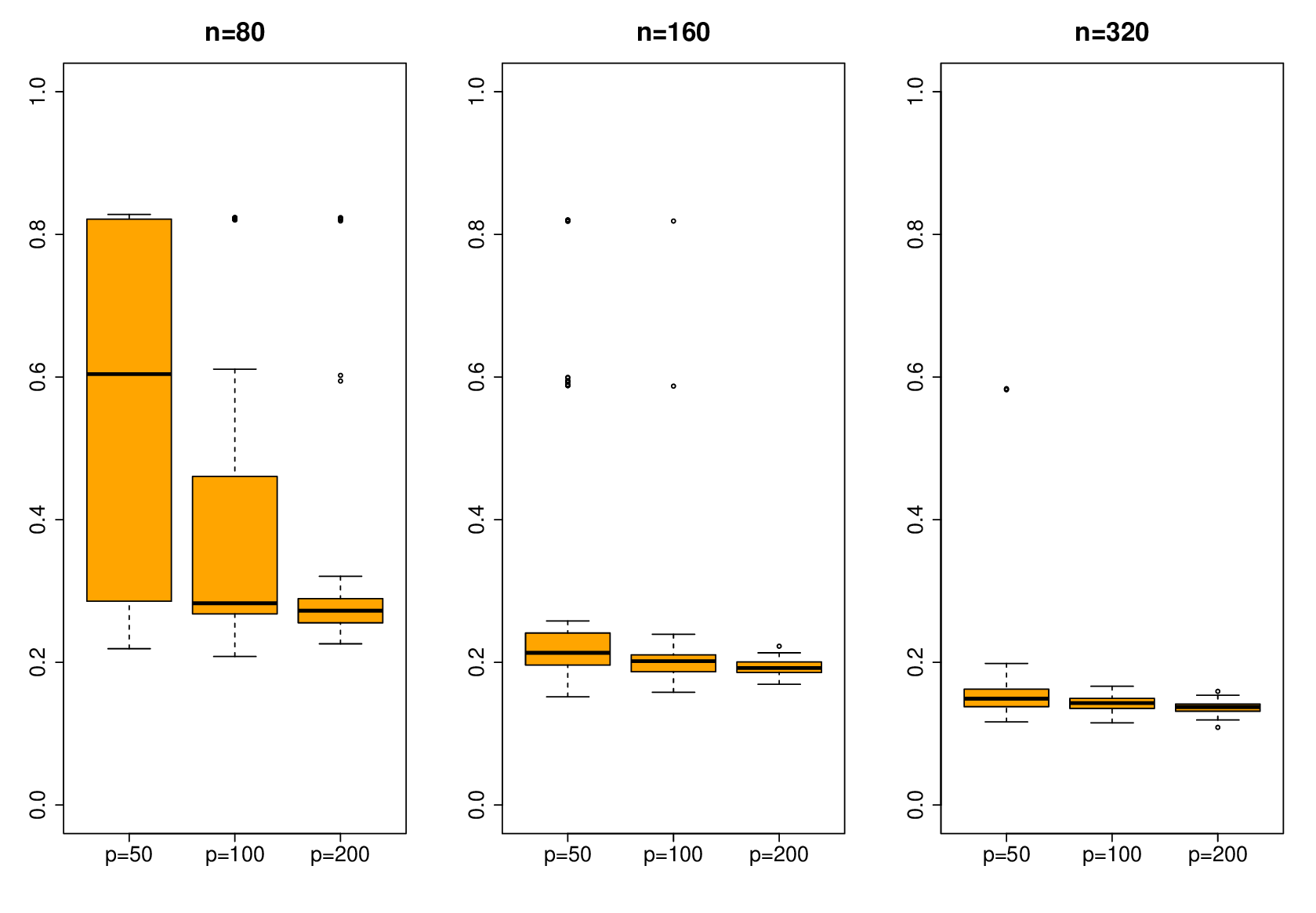, width=10cm}  

	\caption{Boxplot of ${1\over 2} \{ D(\calM(\wh\bA_1),\calM(\bA_1))+
		D(\calM(\wh\bA_2),\calM(\bA_2))\}$ from a simulation with 100 replications.}
	\label{figure_distanceA}
\end{center}
\end{figure}

Fig.\ref{figure_distanceA} depicts the boxplots of the average distance
\[
\frac{1}{2} \{ D(\calM(\wh\bA_1 ), \calM(\bA_1)) + D(\calM(\wh\bA_2 ), \calM(\bA_2)) \}
\]
over 100 replications under different settings.
As expected, the errors in estimating $\calM(\bA_1)$ and $\calM(\bA_2)$ decrease as $n$ increases.
Perhaps more interesting is the phenomenon that the estimation errors do not increase as the number of locations $p$ increases.
Note that the three factors specified in the above model are all strong factors.
According to Proposition \ref{them52}(iii), $D(\calM(\wh \bA_i),
\calM(\bA_i))= O_p(n^{-1/2} + \tau \|\bL\|/p^2)$ when $\delta=0$.
See also Remark \ref{remark2}(i). Fig.\ref{figure_distanceA} also shows that the estimation
errors with $p=100$ are significantly greater than those with $p=200, 400$. This is due
to greater errors in estimating $d$ with smaller $p$; see Table~\ref{table_kriging} below.
Note that Proposition \ref{them52}(iii) assumes $d$ known.

\begin{figure}
\centerline{
\psfig{angle=-90,figure=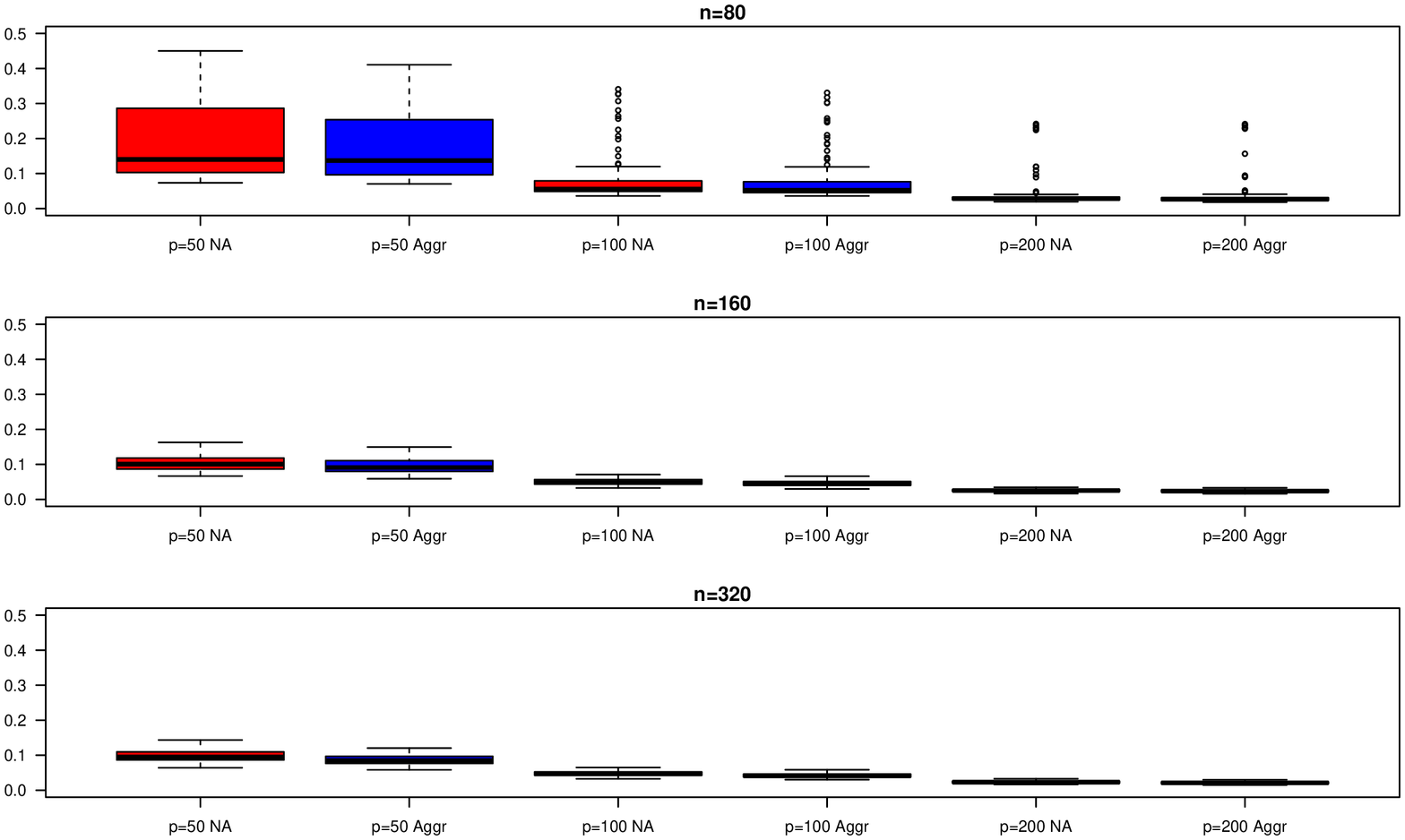, width=13cm } 
}
	\caption{Boxplot of MSE$(\wh \xi )$  (red) and MSE$(\wt \xi )$
(blue) in a simulation with 100 replications.}
	\label{figure_mse_kriging}
\end{figure}

Fig.\ref{figure_mse_kriging} presents the boxplots of
\begin{eqnarray}\label{eqn_mse_xi}
\MSE(\wh \xi ) = {1 \over n p} \sum_{t=1}^n\sum_{j=1}^p \{ \wh
\xi_t(\bs_j) - \xi_t(\bs_j) \}^2, \quad
\MSE(\wt \xi ) = {1 \over n p} \sum_{t=1}^n\sum_{j=1}^p \{ \wt
\xi_t(\bs_j) - \xi_t(\bs_j) \}^2,
\end{eqnarray}
where $\wh
\xi_t(\bs_j)$ and $\wt
\xi_t(\bs_j) $ are defined in, respectively, (\ref{f7}) and (\ref{f8}).
We set $J=100$ for the aggregation estimates $\wt \xi_t(\bs_j) $.
As shown by Theorem \ref{prop2}, $\wt \xi_t( \bs_j)$ always provides
more accurate estimate for $\xi_t( \bs_j)$ than $\wh \xi_t( \bs_j)$.
Furthermore the MSE decreases when either $n$ or $p$ increases.

Note that estimating $\bA_1, \bA_2$ with $\tau>0$ makes use the continuity
of the loading functions $a_i(\cdot)$.  Table \ref{table_cv} lists
the means and the standard errors, over 100 replications,  of MSE$(\wh
\xi )$ with $\wh \xi_t(\bs_j)$
calculated using either $\tau$ selected by
the five-fold cross-validation (i.e. $\tau>0$) or $\tau=0$.
The improvement from using
the continuity is more pronounced when $n$ and $p$ are small.

\begin{table}
	\begin{quote}
		\caption{Means and standard errors (in parentheses) of
MSE$(\wh \xi )$ with $\wh \xi_t(\bs_i)$
			calculated using either $\tau>0$ selected by
			five-fold cross-validation  or  $\tau=0$. } \label{table_cv}
	\end{quote}
	\vspace{-1cm}
	\centering
	\begin{tabular}{rr|cc}
		\hline\hline
		$n$ &  $p$ &  $\tau>0$ 		&  $\tau=0$ \\\hline
		80  &  50  &  0.0941(0.0347) & 0.1139 (0.0429)\\
		160 &  50  &  0.0665(0.0164) & 0.0795 (0.0250)\\
		320 &  50  &  0.0585(0.0076) & 0.0631 (0.0126)\\
		80  &  100 &  0.0243(0.0157) & 0.0279 (0.0183)\\
		160 &  100 &  0.0158(0.0055) & 0.0168 (0.0073)\\
		320 &  100 &  0.0146(0.0011) & 0.0150 (0.0012)\\
		80  &  200 &  0.0056(0.0050) & 0.0064 (0.0058)\\
		160 &  200 &  0.0039(0.0002) & 0.0039 (0.0003)\\
		320 &  200 &  0.0037(0.0002) & 0.0037 (0.0002)\\
		\hline\hline
	\end{tabular}
\end{table}

To illustrate the kriging performance,
with each sample we also draw additional 50 `post-sample' data points at the
 locations randomly drawn from $U[-1, 1]^2$.
For each $t = 1, \cdots, n$,
we calculate the spatial kriging estimate $\wh y_t^r(\cdot)$ in (\ref{f10new}) at
 each of the 50 post-sample locations.
The mean squared predictive error is computed as
\begin{eqnarray}\label{eqn_mspe_kis}
\MSPE(\wh y^r)=  {1 \over 50 n } \sum_{t=1}^n \sum_{\bs_0 \in \calS^* }
\{ \wh y_t^r(\bs_0) - y_t(\bs_0) \}^2,
\end{eqnarray}
where $ \calS^*$ is the set consisting of the 50 post-sample locations.
Similarly, we repeat this exercise for $\wt y_t^r(\cdot)$ in (\ref{f12}).
To check the performance of the kriging in time,
we also generate two post-sample surfaces at times $n+1$ and $n+2$ for each sample.
The mean of square predictive error ($\MSPE$) is calculated as follows.
\begin{eqnarray}
{\rm MSPE}(\wh y^r_{n+\ell})= \frac{1}{p} \sum_{j=1}^p
\{ \wh y_{n+\ell}^r(\bs_j) - y_{n+\ell}(\bs_j) \}^2 , \qquad \ell =1, 2.
\end{eqnarray}
We  repeat the above exercise for the aggregation estimator
$\wt{y}^r_{n+\ell}$ with $J=100$.

\begin{table}
	\caption{Means of $\wh d$, means and standard errors (in parentheses) of 
MSPE for kriging in space and time.}\label{table_kriging}
	\resizebox{\textwidth}{!}{
		\begin{tabular}{rr|c|cc|cccc}
			\hline\hline
			\multicolumn{2}{r}{} & \multicolumn{1}{c}{} & \multicolumn{2}{c}{Kriging over Space} &\multicolumn{4}{c}{Kriging in Time}\\\hline
			$n$ &  $p$ &  $\hat{d}$ & $\MSPE(\hat{y}^r_t)$ & $\MSPE(\tilde{y}^r_t)$ & $\MSPE(\hat{y}^r_{t+1})$ & $\MSPE(\tilde{y}^r_{t+1})$ & $\MSPE(\hat{y}^r_{t+2})$ & $\MSPE(\tilde{y}^r_{t+2})$\\\hline
			 80 & 50  & 2.03 & 1.1893(0.1094) & 1.1763(0.1030) & 1.6300(0.5402) & 1.5660(0.5225) & 1.7856(0.8940) & 1.6876(0.8237)\\
			160 & 50  & 2.76 & 1.1119(0.0553) & 1.1016(0.0467) & 1.3765(0.4160) & 1.3346(0.3952) & 1.4795(0.4749) & 1.4599(0.4754)\\
			320 & 50  & 2.98 & 1.1004(0.0243) & 1.0888(0.0209) & 1.5073(0.5175) & 1.4699(0.4932) & 1.6132(0.7828) & 1.5855(0.7640)\\
			 80 & 100 & 2.62 & 1.0829(0.0765) & 1.0804(0.0735) & 1.5037(0.4135) & 1.4354(0.3904) & 1.8469(0.7127) & 1.7680(0.6564)\\
			160 & 100 & 2.97 & 1.0509(0.0283) & 1.0455(0.0255) & 1.4701(0.4359) & 1.4244(0.4119) & 1.6118(0.5449) & 1.5866(0.5357)\\
			320 & 100 & 3.00 & 1.0462(0.0141) & 1.0412(0.0139) & 1.3541(0.3580) & 1.3290(0.3410) & 1.6301(0.6608) & 1.6137(0.6555)\\
			 80 & 200 & 2.88 & 1.0411(0.0484) & 1.0368(0.0457) & 1.5157(0.4376) & 1.4884(0.4297) & 1.8312(0.7495) & 1.7954(0.7220)\\
			160 & 200 & 3.00 & 1.0238(0.0146) & 1.0221(0.0147) & 1.4471(0.4120) & 1.4326(0.4211) & 1.6841(0.5954) & 1.6721(0.5910)\\
			320 & 200 & 3.00 & 1.0225(0.0122) & 1.0204(0.0121) & 1.4006(0.3285) & 1.3877(0.3299) & 1.5689(0.5047) & 1.5650(0.5111)\\
			\hline\hline
		\end{tabular}
	}
\end{table}

The means and the standard errors  of the MSPE in the 100 replications for each settings are listed in Table \ref{table_kriging}.
In general MSPE decreases as $n$ increases. For the kriging over space, MSPE also decreases
as $p$ increases. See also Theorem \ref{th3}, noting $\delta =0$ when all the factors
are strong.
MSPEs of the kriging over space are smaller than those of the kriging in time.
This is understandable from comparing Theorem \ref{th3} and Theorem \ref{th4}.
The aggregated kriging always outperforms the non-aggregate counterparts.
Last but not least,
the ratio estimator (\ref{c5}) for $d$ works well for reasonably large $n$ and $p$.

\subsection{Real Data Analysis}

We illustrate the proposed methods with 
the monthly temperature records (in Celsius) at the 128 monitoring stations
in China from January 1970 to December 2000.
All series are of the length $n=372$.
For each series, we remove the annually seasonal component by subtracting the average temperature of the same months.
The distance among the stations are calculated as the great circle distance based on their longitudes and latitudes.

For kriging over space, we randomly select $p=78$ stations for
estimation, and predict the values at the other 50 stations.
The mean squared predictive error for the non-aggregation estimates (\ref{f10}) are calculated as follows.
\[
{\rm MSPE}(\wh y^r)=  {1 \over 50 \times 372 } \sum_{t=1}^{372} \sum_{\bs_0 \in \calS^* }
\big\{ \wh y_t^r(\bs_0) - y_t(\bs_0)
 \big\}^2.
\]
We also apply the aggregation (with $J=100$) estimator $\wt y_t(\cdot)$ in
(\ref{f12}) to improve the kriging accuracy.
To avoid the sampling bias in selecting stations,
we replicate this exercise 100 times via randomly
dividing the 128 stations into two sets of sizes 78 and 50.
The estimated $d$-values are equal to 1 in the 98 replications, and are 2 in the
two other replications.
The means of MSPE over the 100 replications for $\hat y^r$ and $\wt y^r$ are 0.7787 and 0.7718,
and the corresponding standard errors are 0.0335 and 0.0444, respectively.
In the training step,
the average MSPE of cross-validation are 0.2407 with optimal $\tau$,
where $\tau>0$, and 0.2493 with $\tau$ equals to zero.
Among all 100 replications, the optimal $\tau$'s are larger than zero for 93 times.

For kriging in time,
we consider one-step-ahead and two-step-ahead post-sample prediction
(with $j_0=6$) for all the 128 locations in each  of the last 24 months in the data set.
The corresponding mean squared predictive error at each step is defined as
\[
{\rm MSPE}(\wh y^r_{n+\ell})= {1 \over 128} \sum_{j=1}^{128}
\big\{ \wh y_{n+\ell}^r(\bs_j) - y_{n+\ell}(\bs_j) \big\}^2, \qquad \ell=1, 2.
\]
We also apply the aggregation estimator $\wt y^r_{n+\ell}(\cdot)$ with $J=100$.
The means and standard errors of ${\rm MSPE}(\wh y^r_{n+\ell})$ over the last 24 months
is 1.7338 and 1.2581 for $\ell =1$, while 1.8814 and 1.4680 for $\ell=2$.
On the other side, the means and standard errors of ${\rm MSPE}(\wt y^r_{n+\ell})$
are 1.7303 and 1.2583 for $\ell =1$, 1.8802 and 1.4673 for $\ell=2$, respectively.
As we expected, the one-step-ahead prediction is more accurate than the two-step-ahead prediction.

Overall the kriging in space is more accurate than those in time.
The aggregation via random partitioning of locations improves the
prediction, though the improvement is not substantial in this example.

\bigskip

\noindent
{\bf Acknowledgements}. We thank Professor Noel Cressie for helpful comments and
suggestions.

\section*{References}
\begin{description}
\begin{singlespace}
\vspace{-3mm}
\item
Banerjee, S., Gelfand, A., Finley, A. O. and Sang, H. (2008). Gaussian
predictive process models for large spatial
data sets. \JRSSB, {\bf 70}, 825-848.

\vspace{-3mm}
\item
Bathia, N., Yao, Q. and Ziegelmann, F. (2010).
Identifying the finite dimensionality of curve time series.
\AS, {\bf 38}, 3352-3386.

\vspace{-3mm}
\item
Breiman, L. (1996). Bagging predictors.
{\sl Machine Learning}, {\bf 24}, 123-140.

\vspace{-3mm}
\item
Castruccio, S. and Stein, M. L. (2013).
Global space-time models for climate ensembles.
{\sl Annals of Applied Statistics}, \textbf{7}, 1593-1611.

\vspace{-3mm}
\item
Chang, J., Guo, B. and Yao, Q. (2015).
High dimensional stochastic regression with latent factors,
endogeneity and nonlinearity. \JOE, {\bf 189}, 297-312.

\vspace{-3mm}
\item
Cressie, N. and Johannesson, G. (2008).
Fixed rank kriging for very large spatial data sets. \JRSSB,
\textbf{70}, 209-226.

\vspace{-3mm}
\item
Cressie, N., Shi, T. and Kang, E.L. (2010). Fixed rank filtering for spatio-temporal
data. {\sl Journal
of Computational and Graphical Statistics}, {\bf 19}, 724-745.

\vspace{-3mm}
\item
Cressie, N. and Wikle, C. K. (2011). {\sl Statistics for Spatio-Temporal Data}.
Wiley, Hoboken.

\vspace{-3mm}
 \item
 Fan, J. and Gijbels, I. (1996). {\sl Local Polynomial Modelling and Its Applications}.
 Chapman and Hall, London.



\vspace{-3mm}
\item
Finley, A., Sang, H., Banerjee, S. and Gelfand, A. (2009). Improving the
performance of predictive process modeling
for large datasets. {\sl Computational Statistics and Data Analysis}, {\bf 53}, 2873-2884.

\vspace{-3mm}
\item
Gneiting, T. (2002). Compactly supported correlation functions.
{\sl Journal of Multivariate Analysis}, {\bf 83}, 493-508.

\vspace{-3mm}
\item
 Golub, G. and Van Loan, C. (1996). {\sl Matrix Computations (3rd edition)}. John Hopkins University Press.

\vspace{-3mm}
\item
Guinness, J. and Stein, M. L. (2013).
Interpolation of nonstationary high frequency spatial-temporal
temperature data. {\sl Annals of Applied Statistics}, \textbf{7}, 1684-1708.

\vspace{-3mm}
\item Hall, P., Fisher, N. I., and Hoffmann, B. (1994). On the
Nonparametric Estimation of Covariance Functions.  \AS, \textbf{22}, 2115-2134.

\vspace{-3mm}
\item
Hastie, T., Tibshirani, R. and Friedman, J. (2009). {\sl The Elements of Statistical Learning}.
Springer, New York.

\vspace{-3mm}
\item
Higdon, D. (2002). Space and space-time modeling using process
convolutions. In {\sl Quantitative Methods for Current
Environmental Issues} (eds C. W. Anderson, V. Barnett, P. C. Chatwin and
A. H. El-Shaarawi), pp. 37-54.  London: Springer.

\vspace{-3mm}
\item Jun, M. and Stein, M. L. (2007). An approach to producing
space-time covariance functions on spheres. {\sl Technometrics}, \textbf{49}, 468-479.

\vspace{-3mm}
\item
Kammann, E. E. and Wand, M. P. (2003). Geoadditive models. {\sl Applied
Statistics}, {\bf 52},
1-18.

\vspace{-3mm}
\item
Katzfuss, M. and  Cressie, N. (2011).
Spatio-temporal smoothing and EM estimation for massive remote-sensing data sets.
\JTSA, \textbf{32}, 430-446.

\vspace{-3mm}
\item
Kaufman, C., Schervish, M. and Nychka, D. (2008). Covariance tapering for
likelihood-based estimation in large
spatial data sets. \JASA, {\bf 103}, 1545-1555.

\vspace{-3mm}
\item
Lam, C. and Yao, Q. (2012).
Factor modelling for high-dimensional time series: inference
for the number of factors.
\AS, {\bf 40}, 694-726

\vspace{-3mm}
\item
Lam, C., Yao, Q. and Bathia, N. (2011). Estimation for latent factors for
high-dimensional time series. {\sl Biometrika}, \textbf{98}, 901-918.

\vspace{-3mm}
\item Li, B., Genton, M. G. and Sherman, M. (2007). A nonparametric
assessment of properties of space-time covariance functions. \JASA,
\textbf{102}, 736-744.

\vspace{-3mm}
\item
 Lin, Z. and Lu, C. (1996). \textsl{Limit Theory on Mixing Dependent Random Variables}. Kluwer Academic Publishers, New York.

\vspace{-3mm}
\item
Mercer, J. (1909). Functions of positive and negative type and their
connection with the theory of integral equations. {\sl Philosophical
Transactions of the Royal Society A},  {\bf 209},  415-446.

\vspace{-3mm}
\item
Sang, H. and Huang J. Z. (2012). A full-scale approximation of covariance
functions for large spatial data sets. \JRSSB, {\bf 74}, 111-132.

\vspace{-3mm}
\item
Smith, R. L., Kolenikov, S. and Cox, L. H. (2003).
Spatiotemporal modelling of PM$_{\rm 2.5}$ data with missing values.
{\sl Journal of Geophysical Research}, {\bf 108}, No.D24, {\footnotesize
DOI:10.1029/2002JD002914}.

\vspace{-3mm}
\item Stein, M. (2008). A modeling approach for large spatial data sets.
{\sl Journal of the Korean Statistical Society}, \textbf{37}, 3-10.

\vspace{-3mm}
\item Tzeng, S.L. and Huang, H.C. (2018). Resolution adaptive fixed rank kriging.
{\sl Technometrics}, to appear.

\vspace{-3mm}
\item Wang, W.T. and Huang, H.C. (2017). Regularized principal component
analysis for spatial data. \JCGS, {\bf 26}, 14-25.

\vspace{-3mm}
\item
Wikle, C. and Cressie, N. (1999). A dimension-reduced approach to
space-time Kalman filtering. {\sl Biometrika}, {\bf 86}, 815-829.

\vspace{-3mm}
\item
Zhang, B., Sang, H., Huang, J. Z. (2015). Full-scale approximations of spatio-temporal
covariance models for large datasets. {\sl Statistica Sinica}, \textbf{25}, 99-114.

\vspace{-3mm}
\item
Zhang, R., Robinson, P. and Yao, Q. (2018). Identifying cointegration by eigenanalysis.
{\sl Available at}
{\tt arXiv:1505.00821}.

\vspace{-3mm}
\item
Zhu, H., Fan, J. and Kong, L. (2014). Spatially varying coefficient model
for neuroimaging data with jump discontinuities. \JASA,  \textbf{109}, 1084-1098.

\end{singlespace}
\end{description}


\newpage

\setcounter{page}{1}

\section*{\normalsize Supplementary document of ``Krigings over space and time based on latent
low-dimensional structures''} 
\centerline{\sc Appendix: Technical proofs}

\setcounter{equation}{0}
\renewcommand{\theequation}{A.\arabic{equation}}

{\bf Proof of Proposition \ref{prop1}}. The first part of the proposition can be proved in
the same manner as Proposition 1 of Bathia \etal (2010), which is omitted.
To prove the second part,
 it follows (\ref{b10}) and (\ref{b10n})  that any eigenfunction of $\Sigma_0$ must be
the linear combination of $a_1, \cdots, a_d$, i.e. $\varphi_i (\bs) = \sum_j \ga_{ij} a_j(\bs)$.
Now it follows from (\ref{b9}) and (\ref{b10n}) that
\begin{align*}
\Sigma_0\circ \varphi_i (\bs) &=
\sum_{k,\ell, j} \sigma_{k \ell} \ga_{ij} a_k(\bs) \inner{a_\ell}{a_j}
= \sum_{k,j} \sigma_{k j} \ga_{ij}  a_k(\bs)
= \sum_k \la_i \ga_{ik} a_k(\bs) = \la_i  \varphi_i (\bs) .
\end{align*}
Since $a_1, \cdots, a_d$ are orthonormal, it must hold that
\begin{equation} \label{x1}
\sum_j \sigma_{k j} \ga_{ij} = \la_i  \ga_{ik} , \qquad k=1, \cdots, d.
\end{equation}
As $\sigma_{k j}$ is the $(k,j)$-th element of matrix $\var(\bx_t)$,
(\ref{x1}) is equivalent to $\var(\bx_t) \bga_i = \la_i \bga_i$, i.e. $\bga_i$ is an eigenvector
of $\var(\bx_t) $ corresponding to the eigenvalue $\la_i$, $i=1, \cdots, d$. Furthermore,
\[
I(i=k) = \inner{\phi_i}{\phi_k} = \sum_{j,\ell} \ga_{ij} \ga_{k\ell} \inner{a_j}{a_\ell}
= \sum_j \ga_{ij} \ga_{kj} = \bga_i' \bga_k.
\]
Thus $\bga_1, \cdots, \bga_d$ are orthogonal. \qed

\vskip3mm
To prove Theorem \ref{prop2}(ii), we first introduce Lemma \ref{lemma3} below.
For the simplicity in presentation, we assume that the $d$ positive eigenvalues
of $\bSigma \bSigma'$, defined in (\ref{f5}), are distinct from each other. Then
both $\bA_1$ and $\bA_2$ are uniquely defined if we line up each of the two
sets of the $d$ orthonormal
eigenvectors (i.e. the columns of $\bA_1$ and $\bA_2$) in the descending
order of their {\red corresponding} eigenvalues, and we require that the first
non-zero element of each those eigenvector to be positive.
See the discussion below (\ref{f5}) above.

Using the same notation as in (\ref{f4}), we
denote by $\wh\bA_{1}^{(j)}, \, \wh\bA_{2}^{(j)}$
the estimated factor loading matrices in (\ref{f4}) with the $j$-th partition,
by $\bSigma^{(j)}$ the covariance matrix in (\ref{f5n}),
and by $\bx_t^{(j)}, \, \bx_{t}^{*(j)}$ the estimated latent factors in (\ref{f6}),
$j=1, \cdots, p_0=p!/(p_1! p_2!)$. Assume that the $d$ positive eigenvalues
of $\bSigma^{(j)} (\bSigma^{(j)})'$ are distinct. Then $\wh\bA_{1}^{(j)}$ and
$\wh\bA_{2}^{(j)}$ can be uniquely defined as above. Now we are ready to state
the lemma.

\begin{lemma} \label{lemma3} Let Condition 1 hold.
 Let the $d$ positive eigenvalues
of $\bSigma^{(j)} (\bSigma^{(j)})'$ be distinct, and Condition 2 hold for $\bx_t^{(j)}$
and $\bx_{t}^{*(j)}$ for all $j=1, \cdots, p_0$.
 Then as $p^{\delta} n^{-1/2}+p^{2\delta-2}\tau\|\bL\|=o(1)$, it holds  that
 \beqn \max_{1\leq j\leq p_0}
\{||\hat\bA_1^{(j)}-\bA_1^{(j)}||+||\hat\bA_2^{(j)}-\bA_2^{(j)}||\}
=O_P(p^{\delta}n^{-1/2}+p^{2\delta-2}\tau\|\bL\|).\nn\eeqn
\end{lemma}

 \noindent {\bf Proof}. Since $\max_{1\leq j\leq p_0}||\hat\bA_2^{(j)}-\bA_2^{(j)}||$ can be shown similarly to $\max_{1\leq j\leq p_0}||\hat\bA_1^{(j)}-\bA_1^{(j)}||$, we only prove $\max_{1\leq j\leq p_0}||\hat\bA_1^{(j)}-\bA_1^{(j)}||$ here. Note that for any $1\leq j\leq p_0$,
 \beqn \label{l3-1}||\wh\bSigma^{(j)} (\wh\bSigma^{(j)})'-\tau\bL-\bSigma^{(j)} (\bSigma^{(j)})'||\leq ||\wh\bSigma^{(j)}-\bSigma^{(j)}||^2+2||\bSigma^{(j)}||\times ||\wh\bSigma^{(j)}- \bSigma^{(j)}||+\tau\|\bL\|.\eeqn
 Since $\bx_t^{(j)}$ satisfies  Condition 2, it follows that $||\bSigma^{(j)}||=O(p^{1-\delta}),$ see Lam \etal (2011). On the other hand, by the mixing condition of $\{\by_t\}$, we have
 \beqn \sup_{j}||\wh\bSigma^{(j)}- \bSigma^{(j)}||^2&=&\sup_{j}||{1\over n}\sum_{t=1}^{n}\{(\by_{t,1}^{(j)}-\bar\by_1^{(j)})(\by_{t, 2}^{(j)}-\bar\by_2^{(j)})'-\mathrm{Cov}(\by_{t,1}^{(j)}, \by_{t,2}^{(j)})\}||^2\nn\\
 &\leq&\sum_{i=1}^{p}\sum_{j=1}^{p}\Big\{{1\over n}\sum_{t=1}^{n}[y_t(\bs_i)-\bar y(\bs_i)][y_t(\bs_j)-\bar y(\bs_j)]-\hbox{Cov}[y_t(\bs_i), y_t(\bs_j)]\Big\}^2 \nn\\
 &=&O_p(p^2/n).\nn\eeqn
 Thus, by (\ref{l3-1}),
 \beqn \label{l3-2}\sup_{j}||\wh\bSigma^{(j)} (\wh\bSigma^{(j)})'-\tau\bL-\bSigma^{(j)} (\bSigma^{(j)})'||=O_p(p^{2-\delta}n^{-1/2}+\tau\|\bL\|).\eeqn
By (\ref{l3-2}) and a  similar argument to Theorem 1 of  Lam \etal (2011), we can show that
\beqn \max_{1\leq j\leq p_0}||\hat\bA_1^{(j)}-\bA_1^{(j)}||=O_P(p^{\delta}n^{-1/2}+p^{2\delta-2}\tau\|\bL\|)\nn\eeqn
and complete the proof of Lemma 3.\qed

\vskip3mm

\noindent{\bf Proof of Theorem \ref{prop2}(ii)}.\quad
 Note that
\beqn\label{pp2} &&\mathrm{E}\Big[{1 \over np} \sum_{t=1}^n\sum_{i=1}^p \big\{ \wh \xi_{t}(\bs_i) - \xi_{t}(\bs_i)\big\}^2\Big{|}\{\xi_{t}(\bs_i), \, y_t(\bs_i)\}\Big]\nn\\
&=&{1 \over npp_0} \sum_{t=1}^n\sum_{j=1}^{p_0}
[\wh\bA_1^{(j)}\wh\bx_t^{(j)} -\bA_1^{(j)}\bx_t^{(j)}]'[\wh\bA_1^{(j)}\wh\bx_t^{(j)} -\bA_1^{(j)}\bx_t^{(j)}]\nn\\
&&+{1 \over npp_0} \sum_{t=1}^n\sum_{j=1}^{p_0}[\wh\bA_2^{(j)}\wh\bx_t^{*(j)} -\bA_2^{(j)}\bx_t^{*(j)}]'[\wh\bA_2^{(l)}\wh\bx_t^{*(j)} -\bA_2^{(j)}\bx_t^{*(j)}]\nn\\
&\equiv&\Sigma_1+\Sigma_2.\nn\eeqn
By Lemma 3, we have
\beqn\Sigma_1&=&{1 \over npp_0} \sum_{t=1}^n\sum_{j=1}^{p_0}
\Big\{[\hat\bA_1^{(j)}(\hat\bA_1^{(j)})'-\bA_1^{(j)}(\bA_1^{(j)})']\bA_1^{(j)}\bx_t^{(j)}
+\hat\bA_1^{(j)}(\hat\bA_1^{(j)})'\ve_{t,1}^{(j)}\Big\}'\nn\\
&&\Big\{[\hat\bA_1^{(j)}(\hat\bA_1^{(j)})'-\bA_1^{(j)}(\bA_1^{(j)})']\bA_1^{(j)}\bx_t^{(j)}
+\hat\bA_1^{(j)}(\hat\bA_1^{(j)})'\ve_{t,1}^{(j)}\Big\}' \nn\\
&=&{1 \over npp_0} \sum_{t=1}^n\sum_{j=1}^{p_0}(\bx_t^{(j)})'(\hat\bA_1^{(j)})'[\hat\bA_1^{(j)}
(\hat\bA_1^{(j)})'-\bA_1^{(j)}(\bA_1^{(j)})']'
[\hat\bA_1^{(j)}(\hat\bA_1^{(j)})'-\bA_1^{(j)}(\bA_1^{(j)})']
\bA_1^{(j)}\bx_t^{(j)}\nn\\
&&+{1 \over npp_0} \sum_{t=1}^n\sum_{j=1}^{p_0}(\bx_t^{(j)})'(\bA_1^{(j)})'[\hat\bA_1^{(j)}(\hat\bA_1^{(j)})'-\bA_1^{(j)}
(\bA_1^{(j)})']'
\hat\bA_1^{(j)}(\hat\bA_1^{(j)})'\ve_{t, 1}^{(j)}\nn\\
&&+{1 \over npp_0} \sum_{t=1}^n\sum_{j=1}^{p_0}(\ve_{t, 1}^{(j)})'\hat\bA_1^{(j)}(\hat\bA_1^{(j)})'[\hat\bA_1^{(j)}(\hat\bA_1^{(j)})'-\bA_1^{(j)}(\bA_1^{(j)})']
\bA_1^{(j)}\bx_t^{(j)}\nn\\
&&+{1 \over npp_0} \sum_{t=1}^n\sum_{j=1}^{p_0}(\ve_{t, 1}^{(j)})'\hat\bA_1^{(j)}(\hat\bA_1^{(j)})'\hat\bA_1^{(j)}(\hat\bA_1^{(j)})'\ve_{t,1}^{(j)}\nn\\
&=&O_p(p^{\delta}/n+p^{(\delta-1)/2}n^{-/2}+p^{\delta-2}\tau\|\bL\|)+{1 \over npp_0} \sum_{t=1}^n\sum_{j=1}^{p_0}(\ve_{t, 1}^{(j)})'\bA_1^{(j)}(\bA_1^{(j)})'\ve_{t,1}^{(j)}.\eeqn
Since $\mathrm{E}\Big({1\over p_0}\sum_{j=1}^{p_0}(\ve_{t, 1}^{(j)})'\bA_1^{(j)}(\bA_1^{(j)})'\ve_{t,1}^{(j)}\Big)^2\leq {1\over p_0}\sum_{j=1}^{p_0}
\mathrm{E}\Big[(\ve_{t, 1}^{(j)})'\bA_1^{(j)}(\bA_1^{(j)})'\ve_{t,1}^{(j)}\Big]^2<\infty,$
it follows from Markov's inequality that
\beqn {1 \over np_0}\sum_{t=1}^n\sum_{j=1}^{p_0}(\ve_{t, 1}^{(j)})'\bA_1^{(j)}(\bA_1^{(j)})'\ve_{t,1}^{(j)}\stackrel{p}{\longrightarrow}
\mathrm{E}\Big[(\ve_{t, 1}^{(j)})'\bA_1^{(j)}(\bA_1^{(j)})'\ve_{t,1}^{(j)}\Big].\nn\eeqn
Thus, by (\ref{pp2}), we have the following two conclusions:
\begin{itemize}
\item[(i)] When $n\rightarrow\infty$, $\Sigma_1=O_p(p^{\delta}/n+p^{(\delta-1)/2}n^{-/2}+p^{\delta-2}\tau\|\bL\|+p^{-1}).$

 \item[(ii)] When $p^{1+\delta}/n+p^{\delta-1}\tau\|\bL\|\rightarrow 0$,
$p\Sigma_1\stackrel{p}{\longrightarrow}\mathrm{E}\Big[(\ve_{t, 1}^{(1)})'\bA_1^{(1)}(\bA_1^{(1)})'\ve_{t,1}^{(1)}\Big].$
\end{itemize}
 Similarly, the above properties hold also for $\Sigma_2.$ Hence,
\beqn \label{pp2-6}\mathrm{E}\Big[{1 \over np} \sum_{t=1}^n\sum_{i=1}^p \big\{ \wh \xi_{t}(\bs_i) - \xi_{t}(\bs_i)\big\}^2\Big{|}\{\xi_{t}(\bs_i), \, y_t(\bs_i)\}\Big]=O_p(p^{\delta}/n+p^{(\delta-1)/2}n^{-/2}+p^{\delta-2}\tau\|\bL\|
+p^{-1}). \, \, \,\eeqn
Further, when $p^{1+\delta}/n+p^{\delta-1}\tau\|\bL\|\rightarrow 0$,
$$\mathrm{E}\Big[{1 \over n} \sum_{t=1}^n\sum_{i=1}^p \big\{ \wh \xi_{t}(\bs_i) - \xi_{t}(\bs_i)\big\}^2\Big{|}\{\xi_{t}(\bs_i), \, y_t(\bs_i)\}\Big]=\mathrm{E}\Big[(\ve_{t, 1}^{(1)})'\bA_1^{(1)}(\bA_1^{(1)})'\ve_{t,1}^{(1)}+(\ve_{t, 2}^{(1)})'\bA_2^{(1)}(\bA_2^{(1)})'\ve_{t,2}^{(1)}\Big]$$
in probability. \qed

\begin{lemma} Let Condition 1  hold and $pn^{-\beta/2}\rightarrow 0$. Then
\beqn \lim_{n\rightarrow\infty} P\left\{\min_{1\leq i\leq p}\lambda_{\min}[n^{-1}\bZ(\bs_i)'\bZ(\bs_i)]\geq c_0/2\right\}= 1.\nn\eeqn
\end{lemma}

 \noindent {\bf Proof}. Let $z_{t}^j(\bs_i), \, j=1, \cdots, m$ be the components of $\bz_t(\bs_i).$ Since $\{\bz_t(\bs_i)\}$ is a stationary $\alpha$-mixing process satisfying Condition 1, by
 Lemma 12.2.2 of Lin and Lu (1996), we have that for any $1\leq j, k \leq m$,
\beqn \mathrm{E}\left|{1\over n}\sum_{t=1}^{n}\{z_t^j(\bs_i)z_t^k(\bs_i)-\mathrm{E}[z_t^j(\bs_i)z_t^k(\bs_i)]\}\right|^\beta
=O(n^{-\beta/2}).\eeqn
Since $m$ is finite, it follows that
\beqn \label{8.4}\mathrm{E}||n^{-1}\bZ(\bs_i)'\bZ(\bs_i)-\hbox{Var}(\bz_t(\bs_i))||_F^\beta
=O(n^{-\beta/2}),\eeqn
where $||\cdot||_F$ denotes the Frobenius norm. Now
suppose that $\min_{1\leq i\leq p}\lambda_{\min}[n^{-1}\bZ(\bs_i)'\bZ(\bs_i)]< c_0/2$.
Since $\min_{1\leq i\leq p}\lambda_{\min}[\hbox{Var}(\bz_t(\bs_i))]> c_0$, and
 $$\hbox{Var}(\bz_t(\bs_i))=
[\hbox{Var}(\bz_t(\bs_i))-n^{-1}\bZ(\bs_i)'\bZ(\bs_i)]+n^{-1}\bZ(\bs_i)'\bZ(\bs_i),$$
 it must hold that
\beqn \max_{1\leq i\leq p}||n^{-1}\bZ(\bs_i)'\bZ(\bs_i)-\hbox{Var}(\bz_t(\bs_i))||_F\geq c_0/2.\eeqn
However, by (\ref{8.4}), it follows that
\beqn P\{\max_{1\leq i\leq p}||n^{-1}\bZ(\bs_i)'\bZ(\bs_i)-\hbox{Var}(\bz_t(\bs_i))||_F\geq c_0/2\}&\leq& \sum_{i=1}^{p}(c_0/2)^{-\beta}
\mathrm{E}||n^{-1}\bz(\bs_i)'\bz(\bs_i)-\hbox{Var}(\bz(\bs_i))||_F^\beta\nn\\
&=&O_p(pn^{-\beta/2})=o(1).\eeqn
This implies that $P\{\min_{1\leq i\leq p}\lambda_{\min}[n^{-1}\bz(\bs_i)'\bz(\bs_i)]< c_0/2\}=o(1)$
and completes the proof of Lemma 4. \qed
\vskip3mm

theorem and Theorem 8.1.10 of Golub and Van Loan (1996) (see also Lemma 3
of Lam \etal (2011)). (ii) can be shown similarly to Theorem 1 of Bathia,
\etal (2010), see also Theorem 1 of Lam and Yao (2012).
%
%
%
%
%
%

\askip

\noindent {\bf Proof for the convergence rate  of $\wh \bbeta(\bs_0)$}.  Let $e_t(\bs)=y_t(\bs)-\bz_t(\bs)'\bbeta(\bs)$ and  $w_i=K_h(\bs_i-\bs_0)/\sum_{i=1}^{p}K_h(\bs_i-\bs_0)$. Then $\be(\bs)=(e_1(\bs), \cdots, e_n(\bs))'$ and
\beqn \wh \bbeta(\bs_0)=\sum_{i=1}^{p}[\bZ(\bs_i)'\bZ(\bs_i)]^{-1}[\bZ(\bs_i)'\be(\bs_i)]
w_i+\sum_{i=1}^{p}\bbeta(\bs_i)
w_i\equiv I_1+I_2.\nn\eeqn
For  any twice differentiable function $g(\bs)=g(s_1, s_2), \, \bs=(s_1, s_2)\in \RR^2$,  define $g_{1\cdot}(\bs)=\partial g(\bs)/\partial s_1,$ $ \, \,
g_{\cdot 2}(\bs)=\partial g(\bs)/\partial s_2,$ $ g_{1 1}(\bs)=\partial^2 g(\bs)/(\partial s_1)^2$ and $g_{2 2}(\bs)=\partial^2 g(\bs)/(\partial s_2)^2$.
Under Conditions 3, 4 and Taylor's expansion, it can be shown that as $p\rightarrow\infty$,
 \beqn\label{8.11} I_2-\bbeta(\bs_0)&=&\sum_{i=1}^{p}(\bbeta(\bs_i)-\bbeta(\bs_0))
w_i\nn\\
&=&{ h^2\over f(\bs_0)}[\bbeta_{1 \cdot}(\bs_0)  f_{1\cdot}(\bs_0)+{1\over 2}f(\bs_0)\bbeta_{1 1}(\bs_0)]\int_{R}\int_R x^2K(x, y)\, dx dy \nn\\
&&+{h^2\over f(\bs_0)}[\bbeta_{\cdot 2}(\bs_0) f_{\cdot 2}(\bs_0)+{1\over 2}f(\bs_0)
\bbeta_{22}(\bs_0)]\int_{R}\int_R y^2K(x, y)\, dx dy+o(h^2). \quad
\eeqn
As for $I_1$, by H\"{o}lder's inequality, it follows that
\beqn \label{8.12}||I_1||&\leq&\Big(\max_{1\leq i\leq p}\|[{\bZ(\bs_i)'\bZ(\bs_i)/ n}]^{-1}\|_F\Big)\sum_{i=1}^{p}\|\bZ(\bs_i)'\be(\bs_i)/n\|
w_i.\, \,\eeqn
By Lemma 4, we have
$\max_{1\leq i\leq p}\lambda_{max}\{(n^{-1}\bZ(\bs_i)'\bZ(\bs_i))^{-1}\}\leq 2/c_0$ holds in probability.  Since
the dimension  of $\bz_t(\bs)$ is fixed, it follows that
\beqn \label{8.13}\max_{1\leq i\leq p}||(n^{-1}\bZ(\bs_i)'\bZ(\bs_i))^{-1}||_F\leq c_1\eeqn
holds in probability for some positive constant $c_1$. On the other hand, it is easy to get that
\beqn \max_{1\leq i\leq p}\mathrm{E}||n^{-1}\bZ(\bs_i)'\be(\bs_i)||=O(n^{-1/2}),\nn\eeqn
hence,
\beqn  \label{8.14}\mathrm{E}\Big[\sum_{i=1}^{p}||n^{-1}\bZ(\bs_i)'\be(\bs_i)||
w_i\Big]=O(n^{-1/2}).\eeqn
It follows from (\ref{8.12}), (\ref{8.13}) and (\ref{8.14}) that
\beqn ||I_1||=O_p(n^{-1/2}).\eeqn
Thus, by (\ref{8.11}) and (\ref{8.14}), we have $|\wh\bbeta(\bs_0)-\bbeta(\bs_0)|=O_p(h^2+n^{-1/2}).$\qed

\askip

\noindent {\bf Proof of Theorem 2}.
 Let $\bx_t^{o}=\bx_tI(\bs_i\in {\cal{S}}_1)+\bx_t^*I(\bs_i\in {\cal{S}}_2).$ Then
\beqn \wh \xi_t(\bs_0)-\xi_t(\bs_0)&=&\sum_{i=1}^{p}(\wh \ba'(\bs_i)\wh \bx_{t}^o-\ba'(\bs_i) \bx_{t}^o)w_i
+\sum_{j=1}^{d}\sum_{i=1}^{p} (a_j(\bs_i)-a_j(\bs_0)) x_{tj}^o w_i\nn\\
&\equiv & J_1+J_2.\eeqn
Similar to (\ref{8.11}), we have
\beqn \left|\sum_{i=1}^{p} [(a_j(\bs_i)-a_j(\bs_0))/||a(\bs_0)||] w_i\right|=O(h^2),\nn\eeqn
which implies that
\beqn \label{8.17}|J_2|= O(h^2)(||\ba(\bs_0)||)\sum_{j=1}^{d} |x_{tj}^o|=O(dh^2\cdot||\ba(\bs_0)||\cdot ||\bx_t^o||)=O_p(h^2),\eeqn
where we use the fact that $||\bx_t^o||=O_p(p^{(1-\delta)/2})$ and $||\ba(\bs_0)||=O(p^{(\delta-1)/2})$, which is followed by
$\lambda_{\min}\{\mathrm{E}(\bx_t^o(\bx_t^o)')\}\asymp p^{1-\delta}$ and
\[(||\ba(\bs_0)||^2) \lambda_{\min}\{\mathrm{E}(\bx_t^o(\bx_t^o)')\}\leq (||\ba(\bs_0)||^2)\mathrm{E}[(\ba'(\bs_0)/||\ba(\bs_0)||)\bx_t^o]^2=\mathrm{E}[\ba'(\bs_0)\bx_t^o]^2
=\mathrm{E}y_t^2(\bs_0)<\infty.\]

 By (iii) of Proposition 3 and the same arguments as in Theorem 2.2 of  Chang \etal (2015), we have that for $i=1, 2$,
\beqn \label{8.18} p^{-1/2}||\wh\bA_i \wh\bx_{t}^o-\bA_i\bx_{t}^o||=O_p(||\wh\bA_i-\bA_i||+n^{-1/2}+p^{-1/2})
=O_p(n^{-1/2}p^\delta+p^{2\delta-2}\tau\|\bL\|+p^{-1/2}),\nn\eeqn
which combining with H\"{o}lder inequality implies that
\beqn \label{8.19}J_1&\leq& \Big\{\sum_{i=1}^{p}[(\wh \ba'(\bs_i)\wh \bx_{t}^o-\ba'(\bs_i) \bx_{t}^o)]^2\Big\}^{1/2}\Big(\sum_{i=1}^{p}w_i^2\Big)^{1/2}\nn\\
&\leq&(\sum_{i=1}^{2}||\wh\bA_i \wh\bx_{t}^o-\bA_i\bx_{t}^o||)\Big(\sum_{i=1}^{p}w_i^2\Big)^{1/2}\nn\\
&=&O_p\{p^{1/2}(n^{-1/2}p^\delta+p^{2\delta-2}\tau\|\bL\|+p^{-1/2})\}O((ph)^{-1/2})\nn\\
&=&
O_p\{p^\delta(nh)^{-1/2}+(ph)^{-1/2}+p^{2\delta-2}h^{-1/2}\tau\|\bL\|\}.\eeqn
Thus, (ii) follows from (\ref{8.17}) and (\ref{8.19}). Similarly, we can show that (\ref{8.19}) holds also for  $\wt\bxi_t(\bs_0)$.

\qed

\vskip3mm

\noindent {\bf Proof of Theorem 3}. For  simplicity, we only show the case with spatial points over ${\cal{S}}_1$, i.e.,
$\by_{t1}=\bA_1 \bx_{t}+\bve_{t,1}.$ For points over ${\cal{S}}_2$ can be shown similarly.
Let $\wh\bSigma_{\ve}(k)={1\over n}\sum_{t=1}^{n-k}(\bve_{t+k, 1}-\bar\bve_1)(\bve_{t, 1}-\bar\bve_1)', \, \wh\bSigma_{x\ve}(k)={1\over n}\sum_{t=1}^{n-k}(\bx_{t+k}-\bar\bx)(\bve_{t,1}-\bar\bve_1)',$  $\wh\bSigma_{\ve x}(k)={1\over n}\sum_{t=1}^{n-k}(\bve_{t+k,1}-\bar\bve_1)(\bx_{t}-\bar\bx)'$ and $\wh\bSigma_{xx}(k)={1\over n}\sum_{t=1}^{n-k}(\bx_{t+k}-\bar\bx)(\bx_{t}-\bar\bx)'$.
It follows that for any $k$,
\beqn \label{8.35}\wh\bSigma_x(k)-\bSigma_x(k)
&=&(\wh\bA'_1-\bA'_1)\bA_1\wh\bSigma_{xx}(k)\bA'_1\wh\bA_1+\wh\bSigma_{xx}(k)\bA'_1(\wh\bA_1-\bA_1)
+(\wh\bSigma_{xx}(k)-\bSigma_x(k))\nn\\
&&
+\wh\bA'_1\wh\bSigma_{\ve}(k)\wh\bA_1
+\wh\bA'_1\bA_1\wh\bSigma_{x\ve }(k)\wh\bA_1
+\wh\bA'_1\wh\bSigma_{\ve x}(k)\bA'_1\wh\bA_1\nn\\
&=: &\sum_{j=1}^{6}L_j.\nn\eeqn

By $||\wh\bA_1-\bA_1||=O_p(n^{-1/2}p^\delta+p^{2\delta-2}\tau\|\bL\|)$, it follows that
\beqn\label{PT5.1} ||L_1||+||L_2||=O(||\wh\bA_1-\bA_1||\cdot ||\wh\bSigma_{xx}(k)||)=O_{p}(n^{-1/2}p+p^{\delta-1}\tau\|\bL\|).\eeqn
By (A.1) of Lam and Yao (2012), we have
 \beqn \label{PT5.2} ||L_3||\leq ||\wh\bSigma_{xx}(k)-\bSigma_x(k)||_F=O(d||\wh\bSigma_{xx}(k)-\bSigma_x(k)||)
=O(p^{1-\delta}n^{-1/2}).\eeqn

It is easy to get that
\beqn ||\wh\bSigma_{x\ve }(k)||=O_p(p^{1-\delta/2}n^{-1/2})= ||\wh\bSigma_{\ve x}(k)||, \, \,
 \hbox{and} \, \, ||\wh\bSigma_{\ve }(k)||=O_p(pn^{-1/2}), \nn\eeqn
 see for example Lemma 2 of  Lam \etal (2011).
Thus,
\beqn \label{PT5.3} ||L_4||+||L_5||+||L_6||=O_{p}(pn^{-1/2}).\eeqn
Combining (\ref{PT5.1}), (\ref{PT5.2}) and (\ref{PT5.3}) yields that for any $0\leq k\leq j_0$,
\beqn\label{PT5.4} ||\wh\bSigma_x(k)-\bSigma_x(k)||=O_p(pn^{-1/2}+p^{\delta-1}\tau\|\bL\|).\eeqn
Thus, by $p^{\delta}n^{-1/2}+p^{2\delta-2}\tau\|\bL\|=o(1)$, we get $pn^{-1/2}+p^{\delta-1}\tau\|\bL\|=o(p^{1-\delta})$ and in probability,
\beqn \label{PT5.5}||\wh\bSigma_x(k)||_{\min}\asymp ||\bSigma_x(k)||_{\min}\asymp p^{1-\delta} \asymp ||\bSigma_x(k)||\asymp ||\wh\bSigma_x(k)||.\eeqn
Since $j_0$ is fixed,  from (\ref{PT5.4}) it follows that
\beqn \label{PT5.6}||\wh \bR_{j_0}-\bR_{j_0}||\asymp ||\bW_{j_0}-\wh\bW_{j_0}||\asymp O_p(pn^{-1/2}+p^{\delta-1}\tau\|\bL\|)\eeqn
and from
(\ref{PT5.5}) it follows  that
\beqn\label{PT5.7} \|\bR_{j_0}\|=O( p^{1-\delta})  \, \, \,\hbox{and}\, \, \, ||\wh\bW_{j_0}^{-1}||\asymp ||\bW_{j_0}^{-1}|| \asymp O_p(p^{\delta-1}).\eeqn
Since
$$||\hat\bx_t-\bx_t||
=||(\hat\bA_1-\bA_1)'\bA\bx_t+(\hat\bA_1-\bA_1)'\bve_{t,1}+\bA_1'\bve_{t,1}||
=O_p(p^{1/2+\delta}n^{-1/2}+p^{2\delta-3/2}\tau\|\bL\|+1)
$$
and $p^{\delta/2}(p^\delta n^{-1/2}+p^{2\delta-2}\tau\|L\|)=o(1)$, it follows that $\|\wh\bX\wh\bX'\|=O_p(p^{1-\delta}).$
Note that
\beqn \wh\bx_{n+j}^r- \bx_{n}(j)
&=&\wh \bR_{j_0}\wh\bW_{j_0}^{-1}\hat\bX- \bR_{j_0}\bW_{j_0}^{-1}\bX\nn\\
&=&(\wh \bR_{j_0}-\bR_{j_0})\wh\bW_{j_0}^{-1}\hat\bX+\bR_{j_0}\wh\bW_{j_0}^{-1}(\bW_{j_0}-\wh\bW_{j_0})
\bW_{j_0}^{-1}\hat\bX+\bR_{j_0}\bW_{j_0}^{-1}(\wh\bX-\bX).\nn\eeqn
By (\ref{PT5.6}) and (\ref{PT5.7}), we have
\beqn ||(\wh \bR_{j_0}-\bR_{j_0})\wh\bW_{j_0}^{-1}\hat\bX||^2
&=&O(||\wh \bR_{j_0}-\bR_{j_0}||^2\cdot||\wh\bW_{j_0}^{-1}||^2\cdot ||\hat\bX\hat\bX'||)=O_p(p^{1+\delta}n^{-1}).\eeqn
Similarly,
\beqn ||\bR_{j_0}\wh\bW_{j_0}^{-1}(\bW_{j_0}-\wh\bW_{j_0})\bW_{j_0}^{-1}\hat\bX||^2
&=&O(||\bR_{j_0}||^2\cdot ||\wh\bW_{j_0}^{-1}||^2\cdot ||\bW_{j_0}-\wh\bW_{j_0}||^2\cdot||\bW_{j_0}^{-1}||^2\cdot \|\hat\bX\hat\bX'\|)\nn\\
&=&O_p(p^{1+\delta}n^{-1}).\eeqn
On the other hand, by (\ref{PT5.7}) and $||\hat\bx_t-\bx_t||=O_p(p^{1/2+\delta}n^{-1/2}+p^{2\delta-3/2}\tau\|\bL\|+1)$, we have
\beqn ||\bR_{j_0}\bW_{j_0}^{-1}(\wh\bX-\bX)||=O_p(p^{1/2+\delta}n^{-1/2}+p^{2\delta-3/2}\tau\|\bL\|+1).\eeqn
Thus,
\[||\wh\bx_{n+j}^r- \bx_{n}(j)||=O_p(p^{1/2+\delta}n^{-1/2}+p^{2\delta-3/2}\tau\|\bL\|+1)\]
holds and  (a) of Theorem  3 is proved.

As for Conclusion (b), by Conclusion (a) and (iii) of Proposition 3, we have
\beqn ||\wh\by_{n+j}^r- \by_{n}(j)||
&=&|| \wh\bA\wh\bx_{n+j}^r-\bA \bx_{n}(j)||\nn\\
&\leq&||(\wh\bA-\bA)\bx_{n}(j)||+||\wh\bA(\wh\bx_{n+j}^r-\bx_{n}(j))||\nn\\
&=&O_p(p^\delta n^{-1/2}p^{1/2-\delta/2} +p^{1/2+\delta}n^{-1/2}+p^{2\delta-3/2}\tau\|\bL\|+1)\nn\\
&=&O_p( p^{1/2+\delta}n^{-1/2}+p^{2\delta-3/2}\tau\|\bL\|+1).\nn\eeqn
This gives (b) as desired and completes the proof of
Theorem 3.\qed

\end{document}

%% file: QYalias.tex
 \@addtoreset{equation}{section}
 \renewcommand{\theequation}{\thesection.\arabic{equation}}
\newcommand{\inner}[2]{ \mbox{$ \langle #1, #2 \rangle $} }

\newtheorem{theorem}{Theorem}
\newtheorem{lemma}{Lemma}
\newtheorem{proposition}{Proposition}

\renewcommand{\hat}{\widehat}
\def\singlespace{\def\baselinestretch{1}\@normalsize}

\def\wh{\widehat}
\def\wt{\widetilde}
\def\askip{\vspace{0.1in}}

\newcommand{\MSE}{{\rm MSE}}

\newcommand{\MSPE}{{\rm MSPE}}

\newcommand{\cov}{{\rm Cov}}

\newcommand{\etal}{\mbox{\sl et al.\;}}

\newcommand{\eg}{\mbox{\sl e.g.\;}}

\newcommand{\tr}{\mbox{tr}}
\newcommand{\var}{\mbox{Var}}

\def\ga{\gamma}

\def\la{\lambda}

\newcommand{\ve}{{\varepsilon}}

\newcommand{\bA}{{\mathbf A}}
\newcommand{\bB}{{\mathbf B}}

\newcommand{\bG}{{\mathbf G}}
\newcommand{\bH}{{\mathbf H}}
\newcommand{\bI}{{\mathbf I}}

\newcommand{\bL}{{\mathbf L}}
\newcommand{\bM}{{\mathbf M}}
\newcommand{\bQ}{{\mathbf Q}}

\newcommand{\bR}{{\mathbf R}}

\newcommand{\bU}{{\mathbf U}}
\newcommand{\bV}{{\mathbf V}}
\newcommand{\bW}{{\mathbf W}}
\newcommand{\bX}{{\mathbf X}}

\newcommand{\bZ}{{\mathbf Z}}
\newcommand{\ba}{{\mathbf a}}
\newcommand{\bb}{{\mathbf b}}

\newcommand{\be}{{\mathbf e}}

\newcommand{\bs}{{\mathbf s}}
\newcommand{\bu}{{\mathbf u}}
\newcommand{\bv}{{\mathbf v}}

\newcommand{\bx}{{\mathbf x}}
\newcommand{\by}{{\mathbf y}}
\newcommand{\bz}{{\mathbf z}}
\newcommand{\balpha} {\boldsymbol{\alpha}}
\newcommand{\bbeta}  {\boldsymbol{\beta}}
\newcommand{\bfeta}  {\boldsymbol{\eta}}

\newcommand{\bSigma}{\boldsymbol{\Sigma}}

\newcommand{\bgamma}{\boldsymbol{\gamma}}
\newcommand{\bga}{\boldsymbol{\gamma}}
\newcommand{\bve}{\mbox{\boldmath$\varepsilon$}}

\newcommand{\bxi} {\boldsymbol{\xi}}

\newcommand{\bzeta} {\boldsymbol{\zeta}}
\newcommand{\bGamma} {\boldsymbol{\Gamma}}

\newcommand{\bD}{{\mathbf D}}

\newcommand{\qed}{\hfill $\blacksquare$ \vspace{3mm}}

\newcommand{\calM}{{\mathcal M}}

\newcommand{\calS}{{\mathcal S}}

\newcommand{\RR}{\EuScript R}

\def\6bullets{\bullet\bullet\bullet\bullet\bullet\bullet}

\def\JRSSB{{\sl Journal of the Royal Statistical Society}, {\bf B}}

\def\JCGS{{\sl Journal of Computational and Graphical Statistics}}
\def\JASA{{\sl Journal of the American Statistical Association}}

\def\AS{{\sl The Annals of Statistics}}

\def\JOE{{\sl Journal of Econometrics}}
\def\JTSA{{\sl Journal of Time Series Analysis}}